\documentclass[pre,twocolumn,showpacs,superscriptaddress]{revtex4-1}
\usepackage[colorlinks,linkcolor=blue,citecolor=blue,urlcolor=blue]{hyperref}
\usepackage{pdfpages}
\usepackage{amsmath}
\usepackage{graphicx}
\usepackage{color}

\begin{document}
\preprint{My working paper}

\title{Social dilemma alleviated by sharing the gains with immediate neighbors}
\author{Zhi-Xi Wu}
\email[]{eric0724@gmail.com}
\affiliation{Institute of Computational Physics and Complex Systems, Lanzhou University, Lanzhou Gansu, 730000, China}
\author{Han-Xin Yang}
\email[]{hxyang01@gmail.com}
\affiliation{Department of Physics, Fuzhou University, Fuzhou 350108, China}

\date{\today}

\begin{abstract}
We study the evolution of cooperation in the evolutionary spatial prisoner's dilemma game (PDG) and snowdrift game (SG), within which a fraction $\alpha$ of the payoffs of each player gained from direct game interactions is shared equally by the immediate neighbors. The magnitude of the parameter $\alpha$ therefore characterizes the degree of the relatedness among the neighboring players. By means of extensive Monte Carlo simulations as well as an extended mean-field approximation method, we trace the frequency of cooperation in the stationary state. We find that plugging into relatedness can significantly promote the evolution of cooperation in the context of both studied games. Unexpectedly, cooperation can be more readily established in the spatial PDG than that in the spatial SG, given that the degree of relatedness and the cost-to-benefit ratio of mutual cooperation are properly formulated. The relevance of our model with the stakeholder theory is also briefly discussed.
\end{abstract}

\maketitle

\section{Introdcution\label{Intro}}
Cooperation, where one individual incurs a cost to help another, at different levels of organization constitutes the fundamental building block of the natural world as well as our human society~\cite{Axelrod2006book,Nowak2006science}. According to the fundamental principles of Darwinian evolution all individuals should act selfishly in order to maximize their fitness, thereby ensuring better survival rates. Thus, the emergence and sustainment of cooperation in a competitive world is a conundrum~\cite{Smith1973nature,Axelrod1981science}, and explaining the evolution of cooperation among selfish individuals is a central problem in both biological and social sciences~\cite{Hofbauer1998book,Nowak2006book,Sigmund2010book}. Evolutionary game theory provides a powerful platform to investigate this issue~\cite{Maynard1982book}. Specifically, two simple games, the prisoner's dilemma and the snowdrift game, as the classical paradigms for studying the evolution of cooperative behavior, have drawn much attention from scientific communities~\cite{Kummerli2007prsb}.

The prisoner's dilemma game (PDG) describes the pairwise interactions of individuals with two behavioral options (or strategies)~\cite{Axelrod2006book}: two players must simultaneously decide whether to cooperate or to defect. A cooperator pays a cost $c$ for the coplayer to yield a benefit $b$, irrespective of the strategies the opponent adopts. A defector pays no cost and does not distribute any benefits. Thus, for mutual cooperation both players receive the reward $R=b-c$, whereas for mutual defection both players get just the punishment $P=0$. With unilateral cooperation, the defector yields the highest payoff, the temptation $T=b$, at the expense of the cooperator bearing the cost $S=-c$, known as the sucker's payoff. In an unstructured population, where all individuals are equally likely to interact with each other, defectors have a higher average payoff than unconditional cooperators. Therefore, natural selection increases the relative abundance of defectors and drives cooperators to extinction~\cite{Hofbauer1998book}. Though defection is the evolutionary stable strategy, all individuals would be better off if they all cooperated, hence the dilemma.

The snowdrift game (SG) can be vividly illustrated by a situation where two drivers are caught in a blizzard and blocked on either side of a snowdrift~\cite{Hauert2004nature}. Each driver has two choices: either getting out of the car to shovel (cooperate: $C$) or staying in the car (defect: $D$). If the snowdrift is cleared, both drivers get a benefit $b$ of getting home. There incurs a cost $c$ for the labor of shoveling, with $b>c>0$. Consequently, if both drivers choose $C$, then they both gain benefit $b$ of getting back home while sharing labor $c$ of shoveling; i.e., both get payoff $b-c/2$. Conversely, if both drivers choose to stay in the car, i.e., $D$, they will still be trapped by the snowdrift and get nothing. If one of the drivers shovels, then both can go home, but the noncooperative driver avoids the labor and gains a perfect benefit $b$, while the cooperative driver gets the benefit $b-c$. According to the replicator dynamics~\cite{Hofbauer1998book}, the equilibrium frequency of cooperators in the snowdrift game is $1-c/(2b-c)$. Hence, in contrast to the PDG, the SG is a simple model for the evolution of cooperation when defection is not an evolutionary stable strategy~\cite{Hofbauer1998book}. Nonetheless, the SG still represents a social dilemma in that the population payoff would be optimal if everybody chose $C$~\cite{Hauert2004nature}.

One of the grand scientific challenges in the research concerns the question of how the outcome of interactions in social dilemma situations can be improved. Over the past few decades, five general rules have been identified that can offset an unfavorable outcome of social dilemmas and lead to the evolution of cooperation, including kin selection~\cite{Hamilton1964jtb1,Hamilton1964jtb}, direct reciprocity~\cite{Robert1971qrb}, indirect reciprocity~\cite{Nowak2005nature}, spatial (or network) reciprocity~\cite{Nowak1992nature,Szabo1998pre,Ohtsuki2006nature}, and group selection~\cite{Wilson1975pnas,Traulsen2006pnas}. Besides, very recent research on the PDG and the SG also highlighted voluntary participation~\cite{Szabo2002prl}, dynamic preferential selection~\cite{Wu2006pre}, degree heterogeneity~\cite{Santos2005prl,Gardenes2007prl,Rong2007pre}, inhomogeneous activity~\cite{Szolnoki2007epl,Szabo2009pre}, dynamical linking~\cite{Pacheco2006prl}, asymmetric interaction and replacement graph~\cite{Ohtsuki2007prl,Wu2007pre}, independent interaction and selection time scales~\cite{Roca2006prl,Wu2009pre,Rong2010pre}, appropriate payoff-aspiration~\cite{Chen2008pre}, social diversity~\cite{Perc2008pre,Yang2009pre}, migration~\cite{Yang2010pre,Jiang2010pre,Roca2011pnas,Chen2012pre}, limiting resources~\cite{Requejo2012prl,Requejo2013pre}, conditional interaction~\cite{Szolnoki2012pre}, coveting environmental fitness~\cite{Wang2011jtb}, and dynamic social networks~\cite{Rand2011pnas,Lee2011prl}, to name but a few, as potent ways to facilitate the evolution of cooperation (see~\cite{Hauert2005ajp,Doebeli2005el,Szabo2007pr,Roca2009plr,Perc2010bio, Nowak2012jtb,Perc2013jrsi} for surveys of this field).

Herein we would like to point out that, to the best of our knowledge, a great number of game models in the literature assume that the involved individuals are selfish and unrelated. Actually, it is the baseline assumption behind the development of game theory in early years. However, no one is self-sufficient and everyone relies on the other for successful survival. In reality, we are more or less, directly or indirectly, related with our surroundings, especially with those who share close genetic proximity~\cite{Dawkins1976book} and geographic proximity~\cite{Nowak1992nature} with us. It is well established that altruism can be apparently favored, provided that altruist and beneficiary are genetically related~\cite{Hamilton1964jtb1,Hamilton1964jtb,Hamilton1963an,Antal2009pnas}. Furthermore, a large body of experimental and field evidence indicates that people genuinely care about each other~\cite{Camerer2003book}; that is, we humans tend to be not only concerned about individual success, but also about that of others~\cite{Fehr2003nature}. Very recently, several insightful works have highlighted the importance of the fraternity~\cite{Szabo2012jtb,Szabo2013jtb}, friendliness~\cite{Thomas2013sr}, or other regarding preference (where the individuals take into consideration not only their own but also their neighbor's utilities in the strategy-decision-making process~\cite{Akcay2009pnas,Taylor2007ev}), in resolving the social dilemma.

Continuing along this line of research, we here also intend to study how relaxing the assumption of unrelatedness will affect the evolution of cooperation in social dilemma games. For this purpose, a payoff sharing mechanism is incorporated into the game dynamics to take into account relatedness of individuals. In particular, a finite fraction of payoffs collected by the individuals from direct game interactions will be shared equally by their nearest neighbors. Our preliminary results show that cooperation can be promoted substantially as the the players are becoming more and more related with each other. Our findings are further demonstrated by using an extended pair-approximation method to theoretically predict the cooperation level. Furthermore, of particular interest is that cooperation can be more readily established in the context of the PDG than that in the SG with appropriate parameters. In the remaining parts of this paper, we will present in detail our main findings and the corresponding explanations.

\section{Model Definition\label{model}}
In this work we study the evolution of cooperation in the PDG and SG on a square lattice with periodic boundary conditions, where initially each player on site $i$ is designated either as a cooperator ($s_i$=C) or defector ($s_i$=D) with equal probability 0.5. Each player engages in pairwise interactions within his von Neumann neighborhood, and self-interactions are excluded. The payoffs reaped by the players are determined by the strategies they adopted, and can be cast into a matrix form. For simplicity but without loss of generality, the elements of the payoff matrix for the PDG are rescaled such that $R=1$, $T=1+r$, $S=-r$, and $P=0$, where $r=c/(b-c)$ denotes the ratio of the costs of cooperation to the net benefits of cooperation, i.e., the cost-to-benefit ratio of cooperative behavior~\cite{Hauert2005ajp,Hauert2004nature,Ohtsuki2006nature}. Accordingly, for the SG, we can set $R=b-c/2$ to be 1 such that the evolutionary behavior of the SG can also be investigated with the single parameter, the cost-to-benefit ratio of mutual cooperation $r=c/(2b-c)$, with which the payoff elements now read as $R=1$, $T=1+r$, $S=1-r$, and $P=0$. The payoff matrices for both the PDG and the SG are summarized in Table~\ref{T1}.

\begin{table}[b]
\caption{Payoff matrices of the two studied evolutionary games: the prisoner's dilemma game (PDG) and the snowdrift game (SG). The two strategies are cooperation (C) and defection (D). The parameter $r$ characterizes the cost-to-benefit ratio of mutual cooperation. Note that $r=c/(b-c)$ in the PDG, and $r=c/(2b-c)$ in the SG. \label{T1}}
\begin{ruledtabular}
\begin{tabular}{cccc}
\textbf{PDG:} &C &D \\
\hline
C & 1 & $-r$\\
D & $1+r$ & 0 \\
\hline\\
\textbf{SG:} &C &D \\
\hline
C & 1 & $1-r$\\
D & $1+r$ & 0 \\
\end{tabular}
\end{ruledtabular}
\end{table}
A full Monte Carlo step consists of all individuals playing games with their immediate neighbors simultaneously, collecting payoffs according to the matrices in the Table~\ref{T1}, preceded by a synchronous strategy-learning process.
In order to introduce relatedness among the players, we assume that a fraction $\alpha$ of the total payoffs collected by each player from direct game interactions will be distributed (shared) equally to his nearest neighbors. As a result, the ultimate payoffs/gains of the player $i$ in a round of game can be written as
\begin{equation}\label{gain}
G_i=(1-\alpha)U_i+\alpha\sum_{j\in\Omega_i}\frac{U_j}{k_j},
\end{equation}
where $U_j$ is the total payoffs harvested by the player $j$ from playing games with his $k_j$ neighbors (on the square lattice $k_j=\mathcal{Z}=4$ for all individuals), and the summation is over all the neighborhood $\Omega_i$ of the focal player $i$. Henceforth, $\alpha$ weighs the magnitude of degree of relatedness among the individuals. The larger the value of $\alpha$, the more tightly the individuals are related to each other. It is worth emphasizing that, according to Eq.~(\ref{gain}), the final amount of gains obtained by each individual is determined not only by the strategies of his immediate neighbors, but also by those of his next-nearest neighbors (whose final gains themselves again depend on the strategies of their own nearest and next-nearest neighbors). By means of this way, all the individuals in the population can be either directly or indirectly, more or less, related to each other. Henceforth, our present model cannot be mapped into a spatial evolutionary game with an effective payoff matrix as has been done in~\cite{Szabo2012jtb}.

Before the start of the next round, each player is allowed to learn from one of his adjacent neighbors and update his strategy. A synchronous strategy-updating is implemented such that the focal player $i$ compares his ultimate gains with that of a randomly chosen neighbor $j$, and adopts the neighbor's strategy with a probability in dependence on the payoff difference~\cite{Hauert2004nature}
\begin{equation}\label{prob}
W(s_i\leftarrow s_j)=\mathrm{max}\left\{0,\frac{G_j-G_i}{\mathcal{Z}\mathcal{G}}\right\},
\end{equation}
where $\mathcal{G}=T-S$ for the PDG and $\mathcal{G}=T-P$ for the SG, ensuring the proper normalization of the probability. The rule of thumb is that only the strategies of individuals with better performance (in terms of the magnitude of their gains) have a chance to be learnt by others.

In what follows, we will show the simulation results carried out in a square lattice population of $256\times256$ individuals, whereby the frequency of cooperation, i.e., the fraction of cooperators in the whole population, was determined within $10^5$ full Monte Carlo steps (MCS) after sufficiently long transients were discarded. Moreover, all the simulation results reported below are averaged over $50$ different realizations in order to assure suitable accuracy. Though the results shown below are obtained for synchronous strategy-updating, we note that no qualitative changes occur if we adopt an asynchronous updating of strategies, or if we use an alternative Fermi-function like rule~\cite{Blume1993geb,Szabo1998pre,Szabo2007pr} for the strategy updating (please see the Supplemental Material~[\onlinecite{abc}] for more details).

\section{Results and Analysis\label{results}}

\begin{figure*}
\includegraphics[width=0.7\linewidth]{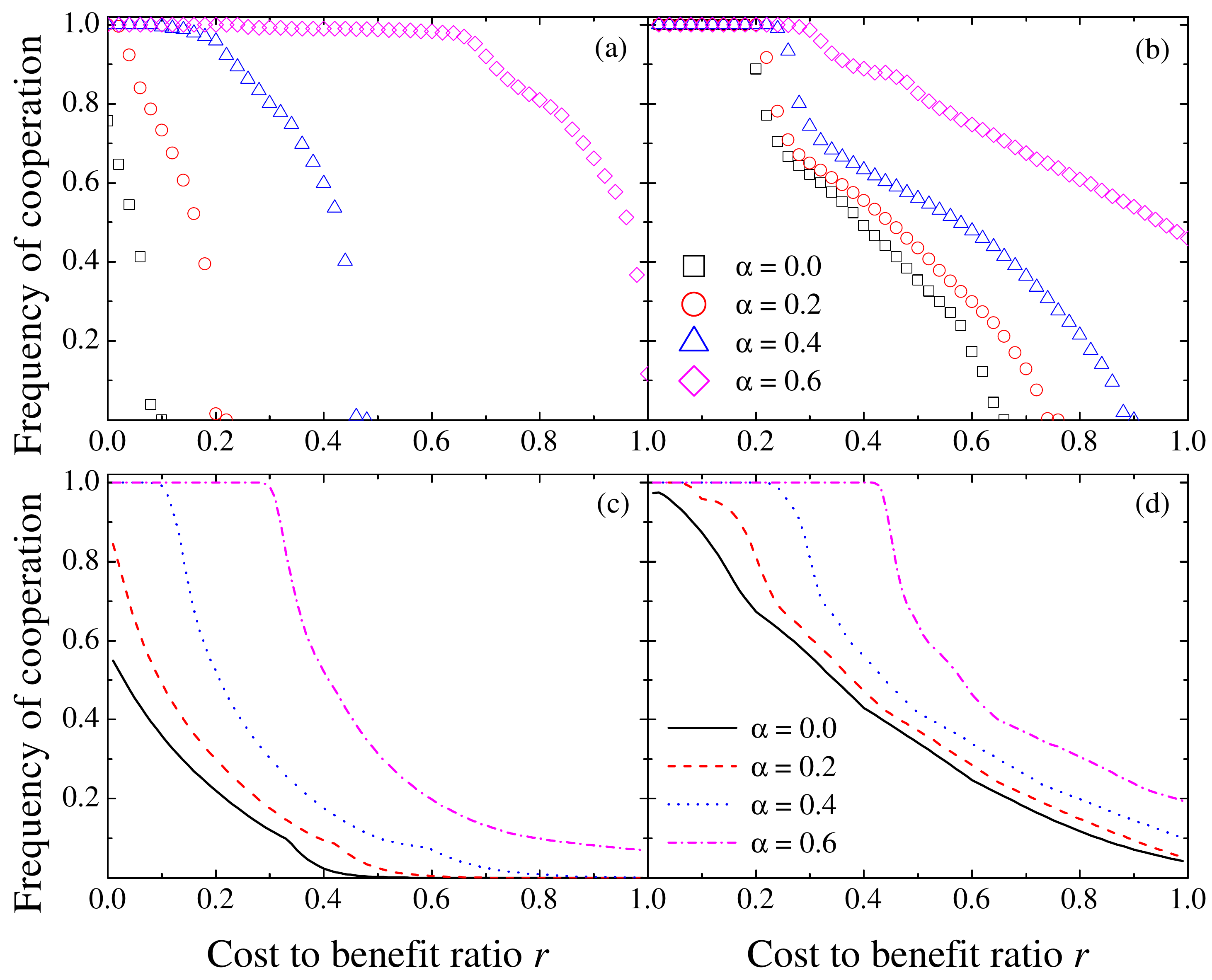}
\caption{(Color online) Frequency of cooperation as a function of the cost-to-benefit ratio (a) $r=c/(b-c)$ in the PDG and (b) $r=c/(2b-c)$ in the SG for different values of $\alpha$, as denoted in the figure legend. The players are located on the sites of a square lattice of size $N=256\times 256$. The curves in (c) and (d) correspond, respectively, to the calculations by the extended pair approximation approach for the PDG and the SG, which correctly predicts the trend, yet significantly overestimates (underestimates) the benefits of population structures for small (large) value of $\alpha$. The error bars in the data points shown in (a) and (b) are smaller than the symbol size, and hence are omitted for clarity. \label{fig1}}
\end{figure*}

First we feature the frequency of cooperation as a function of the cost-to-benefit ratio $r$ for the PDG and the SG for different values of $\alpha$, as shown in Figs.~\ref{fig1}(a) and (b), respectively. We observe that as the players become more and more related to their neighbors (increasing $\alpha$), the frequency of cooperation in the whole population can be greatly promoted under the same cost-to-benefit ratio parameter $r$. Note that for $\alpha=0$ our model reduces to those well-studied classical games, wherein the players are unrelated with each other~\cite{Szabo2007pr,Roca2009plr,Perc2010bio,Nowak2012jtb,Perc2013jrsi}.

Let us first consider the case of the PDG. For the PDG with $\alpha=0$, we note that there exists a critical value of the cost-to-benefit $r_c\approx0.1$ in the PDG, beyond which cooperators cannot survive in the population. Notably, $r_c$ quickly increases to much larger values as $\alpha$ is varied from 0 to 1.0, which means that cooperation is considerably supported (or the social dilemma is heavily alleviated) as the fate of the players become more and more closely correlated. More precisely, $r_c$ is about 0.1 for the standard game $\alpha=0$, while it approaches around 0.2 and 0.46 for $\alpha=0.2$ and $0.4$, respectively. Moreover, for large $\alpha$ there arises a lower critical value of $r$ below which full cooperation can be achieved, and the larger the $\alpha$, the more expanded the region of full cooperation. Particularly, cooperation can still attain a considerable level even if the cost-to-benefit parameter $r=c/(b-c)$ is up to 1.0 when $\alpha=0.6$. For sufficiently large values of $\alpha$, e.g., $\alpha>0.7$, cooperators are able to dominate the whole population for any $r$ in the regime [0,1] (results not shown). The tendency of the frequency of cooperation as a function of $r$ is also correctly predicted by the extended pair-approximation analysis (see the Appendix for more details), as shown in Fig.~\ref{fig1}(c), although it significantly overestimates the extinction threshold of cooperation.

\begin{figure}
\includegraphics[width=\linewidth]{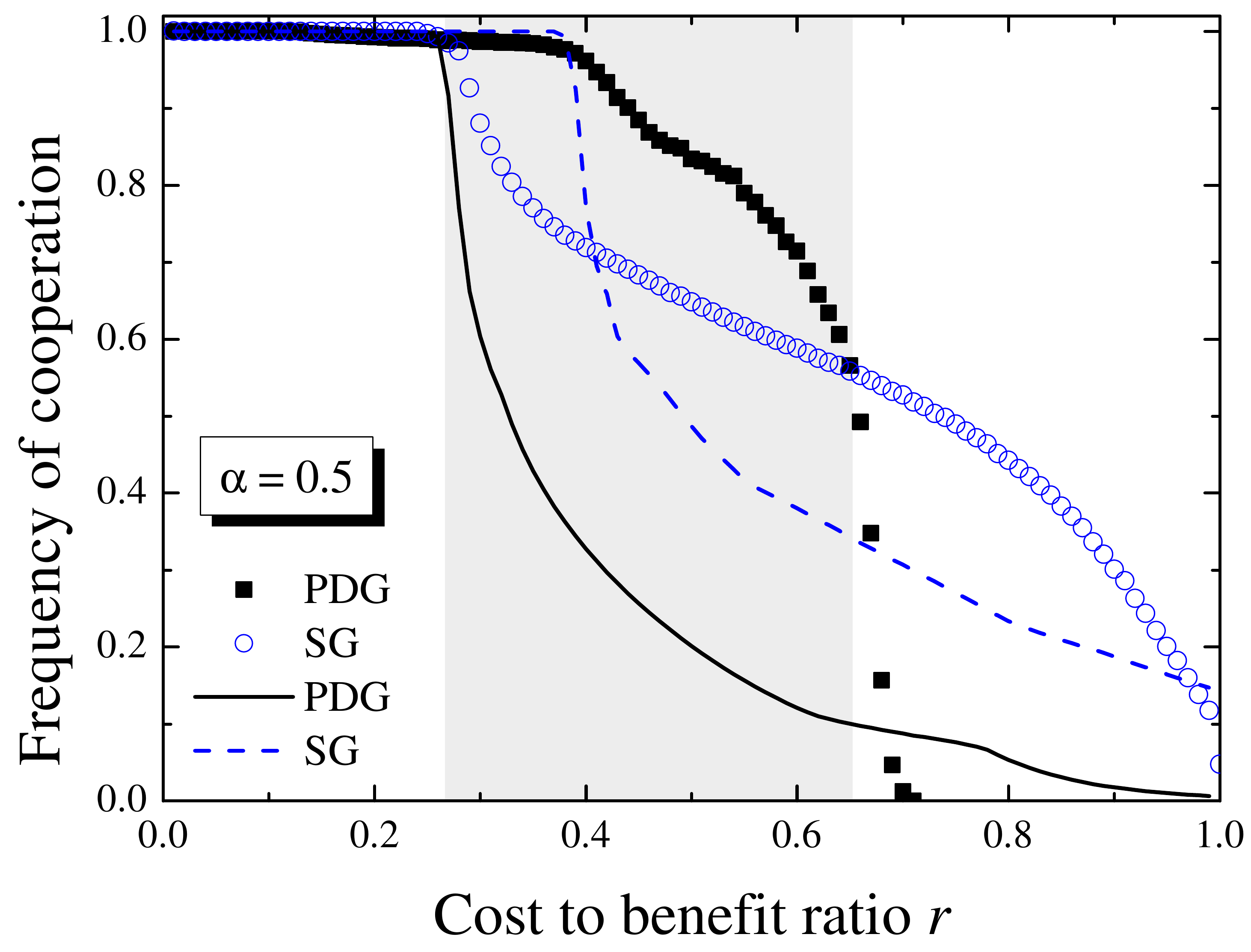}
\caption{(Color online) Frequency of cooperation as a function of the cost-to-benefit ratio parameter $r$ for $\alpha=0.5$ in the square lattice population of size $N=256\times 256$. The symbols and curves correspond to the results obtained by the numerical simulations and the extended pair--approximation, respectively. Note that $r=c/(b-c)$ in the PDG and $r=c/(2b-c)$ in the SG. The shadowed area shows the region where the frequency of cooperation in the PDG is greater than that in the SG with the identical cost-to-benefit ratio of cooperative action. Note that the extended pair approximation fails to predict the crossover behavior of the simulation results, which highlights the vital importance of spatial structure in sustaining and promoting cooperation in the PDG. \label{fig2}}
\end{figure}

Now we turn our attention to the case of the SG. Figure~\ref{fig1}(b) shows equilibrium proportions of cooperators in the lattice population as a function of the cost-to-benefit ratio $r=c/(2b-c)$. As can be observed, in all cases, cooperation is facilitated altogether for any positive values of $\alpha$, which is similar to the scenario in the PDG case. To be more specific, the threshold above which the proportion of cooperators goes to zero increases with increasing $\alpha$, and even disappears for sufficiently large $\alpha$ (note that we only consider the region where $r$ is of physical meaning; i.e., $r\in [0,1]$). In accordance with our simulation results, the extended pair-approximation method qualitatively reflects the role of relatedness in cooperation, but underestimates the resulting cooperation level in general, as indicated by the curves in Fig.~\ref{fig1}(d). It is worthy pointing out that the discrepancy between the simulations and the predictions by the pair-approximation approach is attributed to the fact that the extended pair approximation does not fully take into account the effects of the spatial structures, especially spatial clusters~\cite{Szabo2007pr,Hauert2005ajp,Hauert2004nature,Chen2008preb}.

\begin{figure}
\includegraphics[width=\linewidth]{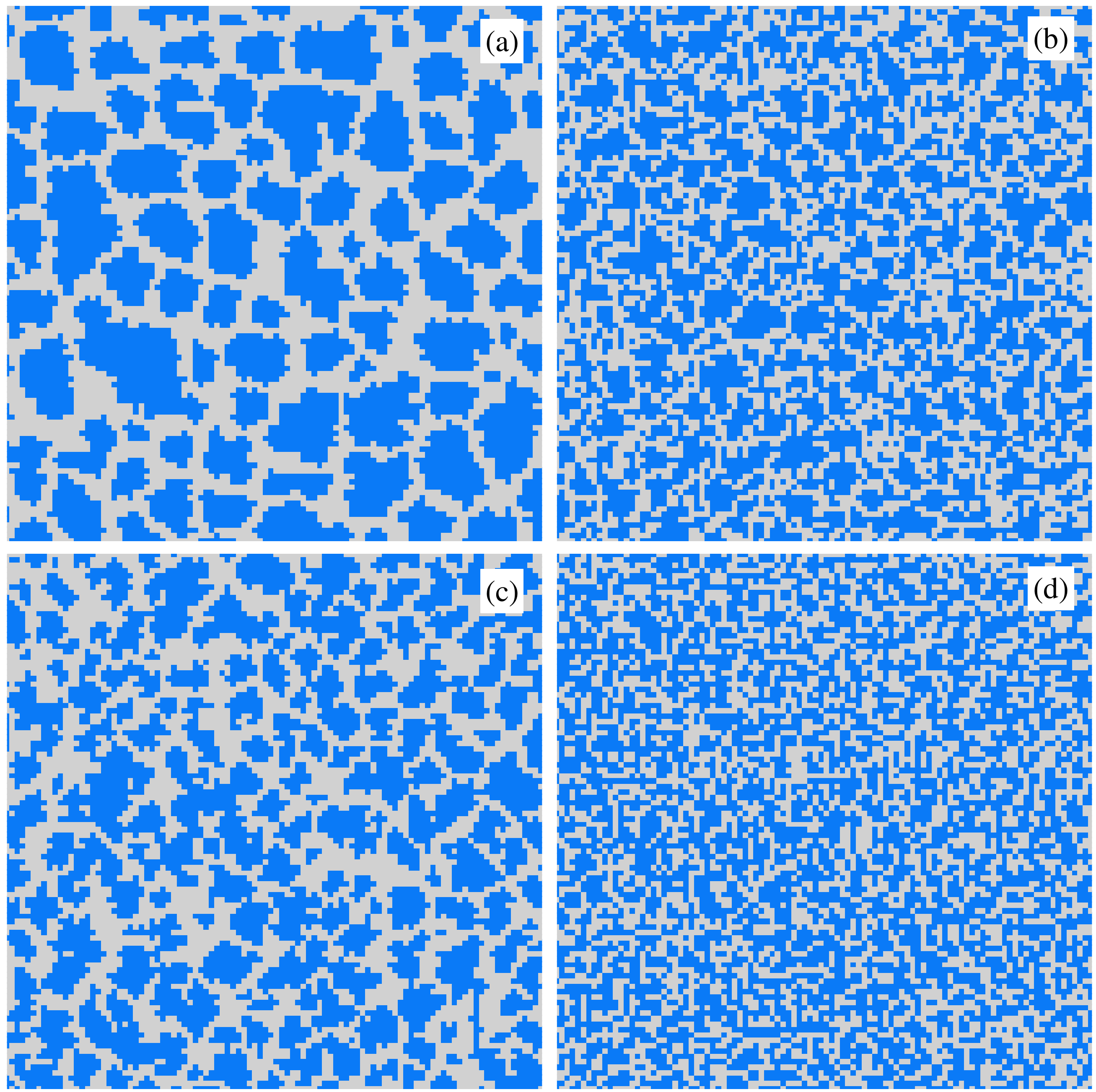}
\includegraphics[width=\linewidth]{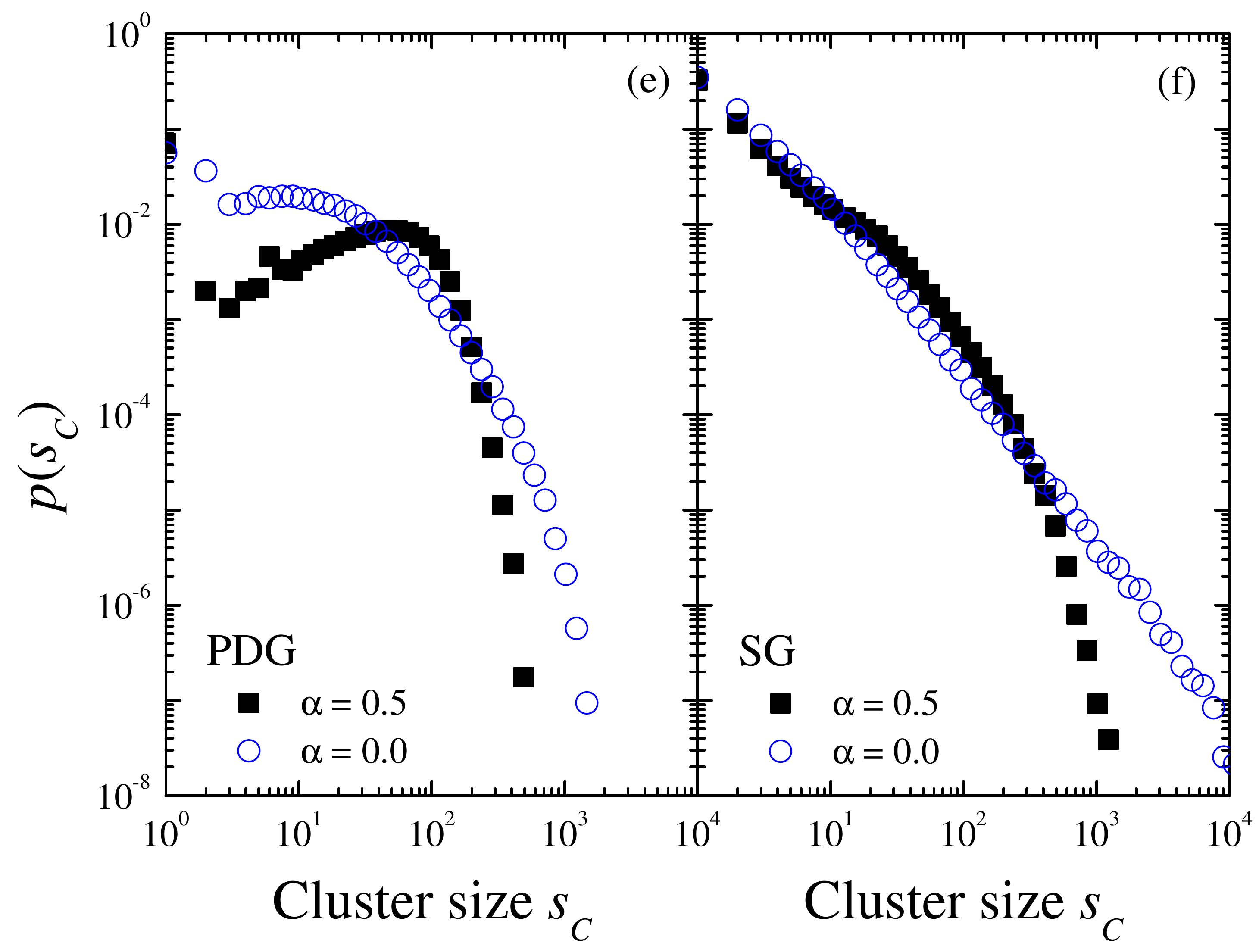}
\caption{(Color online) Panels (a), (b), (c), and (d): Typical snapshots of the strategy configuration of the individuals in the stationary state for the two studied games, where $C$ and $D$ are colored by blue (dark gray) and light gray respectively. The left panels [(a) and (c)] are for the PDG, and the right panels [(b) and (d)] are for the SG. Note that in each case only a $100\times100$ snapshot of spatial configuration of cooperators and defectors of the total $256\times256$ field is shown. We calibrate parameters such that the fraction of cooperators in the steady state are nearly the same in all the considered cases: (a) $\alpha=0.5$, $r=0.65$ (PDG); (b) $\alpha=0.5$, $r=0.65$ (SG); (c) $\alpha=0.0$, $r=0.038$ (PDG); (d) $\alpha=0.0$, $r=0.35$ (SG). With these parametrizations, the frequency of cooperation in the equilibrium is (a) 0.558(3), (b) 0.554(6), (c) 0.563(9), and (d) 0.566(2), respectively. Panels (e) and (f) display the cluster size distribution of cooperators in the equilibrium for the four parameter combinations in the two games (indicated in the panels). The data points shown in (e) and (f) are determined after a sufficiently long transient time ($5\times10^{5}$ MCS), and then averaged over 200 independent runs. The error bars are comparable to the size of the symbols and are omitted for clarity. \label{fig3}}
\end{figure}

The simulation results as well as the predictions by the extended pair-approximation method, summarized in Fig.~\ref{fig1}, explicitly show that increasing the relatedness among the players can strongly facilitate the evolution of cooperation in both the PDG and the SG. It was previously pointed out that spatial structure fails to enhance cooperation in the snowdrift game and actually tends to reduce the proportion of cooperators when the cost-to-benefit is too large~\cite{Hauert2004nature}. Nevertheless, the preliminary results of our model show explicitly that spatial structure favors the evolution of cooperation in the context of the SG, especially when the degree of relatedness among the individuals becomes large enough.

The enhancement of cooperation by strengthening the degree of relatedness among the players can be understood naturally as follows. It is well known that for the PDG in well-mixed populations, cooperators can not outperform defectors and are doomed to extinction for any $r>0$~\cite{Hauert2005ajp}, while in spatially structured populations, cooperators can survive or even thrive by means of forming tight compact clusters to minimize the exploitation by periphery defectors~\cite{Nowak2006science}. The trick lies in the fact that by forming large compact clusters the cooperative individuals can interact with one another more often than they would purely by chance so that they are able to resist the invasion of boundary defectors. Whenever the payoff of an individual is determined not only by the behaviors of his immediate neighbors but also by those of his next-nearest neighbors, i.e., Eq.~(\ref{gain}), the clustering of cooperators reinforces further the interdependence among them, providing a more promising avenue for their thrive. Undoubtedly, such reinforcement mechanism also works for the cooperators in the SG. Furthermore, unlike the case that neighboring cooperators will help each other out, neighboring defectors will just craft their own demise in both the PDG and the SG, hence undermining the contribution of increasing relatedness among them. Taking all together, the presence of relatedness among the individuals is capable of supporting substantially the evolution of cooperation.

According to the replicator dynamics in well-mixed populations~\cite{Hofbauer1998book}, pure defection is the only stable equilibrium strategy in the PDG, while the equilibrium frequency of cooperation in the SG is $1-r$, where $r=c/(2b-c)$ is the cost-to-benefit ratio of mutual cooperation as before. That is to say, in contrast to the PDG, cooperation is already maintained in the well-mixed version of the SG. Likewise, in the spatial versions of the two games, the frequency of cooperation in the SG is always superior to that in the PD under the same cost-to-benefit ratio $r$~\cite{Doebeli2005el,Hauert2005ajp,Szabo2007pr} [also see Figs.~\ref{fig1}(a) and (b)]. Thus, a well established view so far is that cooperation is easier to develop and can more easily persist in the context of the SG than that in the PDG.

However, we find such argument is not always accurate when the relatedness of the neighboring individuals is incorporated into the spatial games, especially when the degree of relatedness is large enough. To get a clear inspection on this point, we plot in Fig.~\ref{fig2} the fraction of cooperators in the equilibrium as a function of $r$ for both the PDG and the SG with $\alpha=0.5$. We observe that for sufficiently high or low $r$, the frequency of cooperation in the SG is greater than that in the PDG, which is in line with the common belief that cooperation can be more easily established in the SG than in the PDG. Very surprisingly, for intermediate values of $r$, the reverse situation occurs. The shadowed area in Fig.~\ref{fig2} shows the region where the frequency of cooperation in the PDG is greater than that in the SG with the identical cost-to-benefit ratio of mutual cooperation.

The ultimate reason for this ``anomalous" phenomenon is that cooperators persist in the two games by means of different manners. In the spatial PDG, cooperators maintain by forming large, compact clusters to reduce exploitation by defectors. By contrast, the cooperators in the spatial SG usually form filament-like structures~\cite{Hauert2004nature}. Given that the total volume is the same, a compact cluster of cooperators will have many fewer boundary defective neighbors than a cluster of cooperators with filament-like structures has. Moreover, as already pointed out by Hauert and Doebeli~\cite{Hauert2004nature}, the emergent dendritic patterns in the SG generate an advantage for defectors, owing to increased exploitation in the fractal-like zone of contact between the two strategies. As such, the presence of relatedness will benefit much more the cooperators in the compact clusters than it does on the cooperators with dendritic structures. Consequently, the combination of localized interactions, relatedness among neighboring individuals, and the different manners of persistence of cooperators makes it possible that cooperation is more readily developed and expanded in the PDG than in the SG for moderate $r$. We want to emphasize that the extended pair approximation fails to predict the crossover behavior of the simulation results, since it neglects most of the spatial correlations (see the Appendix of~\cite{Szabo2007pr} and the Supplemental Information of ~\cite{Hauert2004nature} for more details), which on the other hand highlights the vital importance of spatial structure in sustaining and promoting cooperation in the PDG.

The above argument is further corroborated by inspecting the distribution of the strategy in the equilibrium. In Figs.~\ref{fig3}(a)--(d), several typical snapshots of the spatial configurations of cooperators and defectors after sufficiently long relaxation time are displayed for the PDG and the SG with and without the presence of payoff-sharing mechanism (i.e., relatedness). The two panels in the left (right) column display the results yielded for the PDG (SG). For the sake of comparison, we calibrate parameters such that the fraction of cooperators in the steady state are nearly the same in all the considered cases. In particular, we choose $\alpha=0.5$ and $r=0.65$ in Figs.~\ref{fig3}(a) and(b), $\alpha=0.0$ and $r=0.038$ in Fig.~\ref{fig3}(c), and $\alpha=0.0$ and $r=0.35$ in Fig.~\ref{fig3}(d). With these parametrizations, the frequency of cooperation in the equilibrium is about 0.56 for all the four cases. It is remarkable that with the involvement of payoff-sharing mechanism the cooperators in the PDG are able to expand to form much larger clusters, due to the enhanced reinforcement of reciprocity among the cooperators forming compact structures [Fig.~\ref{fig3}(a)]. In contrast, the expansion of the clusters of cooperators in the SG is not so evident since both cooperators and defectors will benefit from the payoff-sharing mechanism owing to the dendritic-like structures (to be more precise, the more rough surface) formed by the cooperators[Fig.~\ref{fig3}(d)]. The distribution $p(s_C)$ of the cluster size of cooperators $s_C$ in the steady sate further substantiates this point. As illustrated in Fig.~\ref{fig3}(e), there arises a much more visible characteristic cluster size for the case of $\alpha=0.5$ as compared to the case of $\alpha=0.0$ for the PDG, while the shapes of the two distributions in Fig.~\ref{fig3}(f) for the SG deviate from one another in a less significant way. Consequently, the overall results summarized in Figs.~\ref{fig3} show clearly that the presence of relatedness among the players exerts much stronger influence on the stationary strategy distribution in the PDG than that in the SG. As a result, cooperative strategy could be more successful in the context of PDG than that in the SG, provided that the degree of relatedness and the cost-to-benefit ratio of mutual cooperation are properly formulated.

\section{Conclusion and Discussion}
To sum up, we have studied how the presence of relatedness among the players affects the evolution of cooperation in the framework of the prisoner's dilemma game and the snowdrift game. The relatedness is incorporated into the game dynamics by redistributing a fraction $\alpha$ of the payoffs gained by the individuals from direct game interactions to their immediate neighbors. By tuning the parameter $\alpha$, we are able to control the degree of relatedness among the players. The time dependence of the strategy distribution of the individuals is controlled by the imitation of a better performing neighbor. By means of extensive Monte Carlo simulations as well as analytical treatments, we have demonstrated that plugging into relatedness can significantly promote the evolution of cooperation in the context of both games. The larger the fraction of the ultimate payoffs is contributed by the neighboring individuals, the more readily the cooperation can be established and persist in the population. Particularly, for high enough $\alpha$, i.e., when the neighboring individuals are closely related to each other, cooperators can repel defectors completely and take over the whole population. Furthermore, we have shown that, contrary to our common sense, cooperation can more easily expand and be maintained in the context of the PDG than in the SG, provided that the parameter $\alpha$ and the cost-to-benefit ratio $r$ are properly formulated. We explain this by revealing the different organizational patterns of the cooperators in the the two studied games, through which the presence of relatedness among the individuals exerts different effects on the evolution of cooperation in the two games. In particular, the compact clusters formed by the cooperators in the PDG are in favor of cooperators \emph{boosting} each other's success efficiently in the presence of relatedness, while in the SG such reinforcement is not so strikingly evident due to the dendritic-like structures formed by the cooperators.

It is worth making some comparisons between our current work and two closely relevant works by Szab\'{o} \emph{et al.}~\cite{Szabo2012jtb} and Grund \emph{et al.}~\cite{Thomas2013sr}. In these two works, an additional personal feature characterizing the fraternal attitude or friendliness degree of the individuals is introduced to account for the inclination of other-regarding preference of them. The myopic strategy update (myopic best response rule in~\cite{Thomas2013sr}) is assumed for the strategy-updating process: Rather than maximizing their own incomes during the game, the individuals try to maximizing a utility function composed from their own and the co-players' incomes with weight factors $(1-q)$ and $q$ respectively. It was shown that the highest total income is achieved by the society whose members share their income fraternally~\cite{Szabo2012jtb}. Though the role of the tunable parameter $q$ is somewhat similar to the parameter $\alpha$ in Eq.~(\ref{gain}), there are several significant differences between their models and ours. In particular, the utility function in Refs.~\cite{Szabo2012jtb,Thomas2013sr} can be conveniently regarded as a \emph{virtual} payoff, by which the individuals aim to optimize. By contrast, in our model the time dependence of the strategy distribution is governed by a dynamical rule resembling the Darwinian selection and the individuals intend to promote their \emph{actual} payoffs through imitating more successful neighbors. Since the payoff sharing is the rule of game per se in our context, we do not need an extra personal feature to characterize the other-regarding preference of the individuals, and they just take decisions without considering the payoff or utility of others, as in common practices~\cite{Szabo2007pr,Roca2009plr,Perc2010bio,Nowak2012jtb,Perc2013jrsi}. In this sense, our setup of the game is somewhat more simple and neat. Moreover, as we mentioned before, each individual's final payoff in our model is determined by the strategies (or behavior) of his nearest and next-nearest neighbors (whose final incomes in turn depend on their own nearest and next-nearest neighbors). As such, with the payoff-sharing mechanism of Eq.~(\ref{gain}), the fates of the individuals are more involved with each other in our context (the individuals only care about the payoffs of themselves and/or of their direct neighbors in~\cite{Szabo2012jtb,Thomas2013sr}). An alternative view on $\alpha$ in Eq.~(\ref{gain}) is to consider the existence of a \emph{virtual super-organizer} of the games who will always withdraw a fraction $\alpha$ of the payoffs obtained by each individual from direct game interactions, and distribute it equally to all the participates with direct interactions with the focal player. Though we do not need the super-organizer at all in explaining the effectiveness in promoting cooperation of the proposed payoff-sharing mechanism, such perspective is still meaningful: it will be of particular convenience to test our current idea in online experiments with human subjects~\cite{Rand2011pnas}, where the payoff-sharing process can be easily done by the computer.

Associating the players' ultimate payoffs with their neighbors' payoffs is equivalent to letting them sit in the same boat such that they share a common destiny. In the jargon of economics, the players and their neighbors are becoming  ``stakeholders" in the system. According to the stakeholder theory within the economics literature~\cite{Thomas1995}, trusting and cooperative relationships among stakeholders turn out to be more productive, and entities engaging in trusting, trustworthy, and cooperative behaviors will have a significant competitive advantage over those that do not use such criteria. Mapping the level of cooperation onto the extent of productivity, the presented results of our model are in line with the findings in~\cite{Thomas1995}. Due to the simplicity and fundamental character of the model proposed by us, we hope that it might serve as a starting point to inspire future research work aimed at investigating the role of relatedness in the evolution of collective cooperation in real human experiments.

\begin{acknowledgments}
This work was partially supported by the National Natural Science Foundation of China under Grants No. 11005051 and No. 11135001, the Natural Science Foundation of Fujian Province of China under Grant No. 2013J05007, and the Research Foundation of Fuzhou University under Grant No. 0110-600607.
\end{acknowledgments}

\section*{Appendix: Extended pair-approximation method}

\begin{figure}
\includegraphics[width=0.8\linewidth]{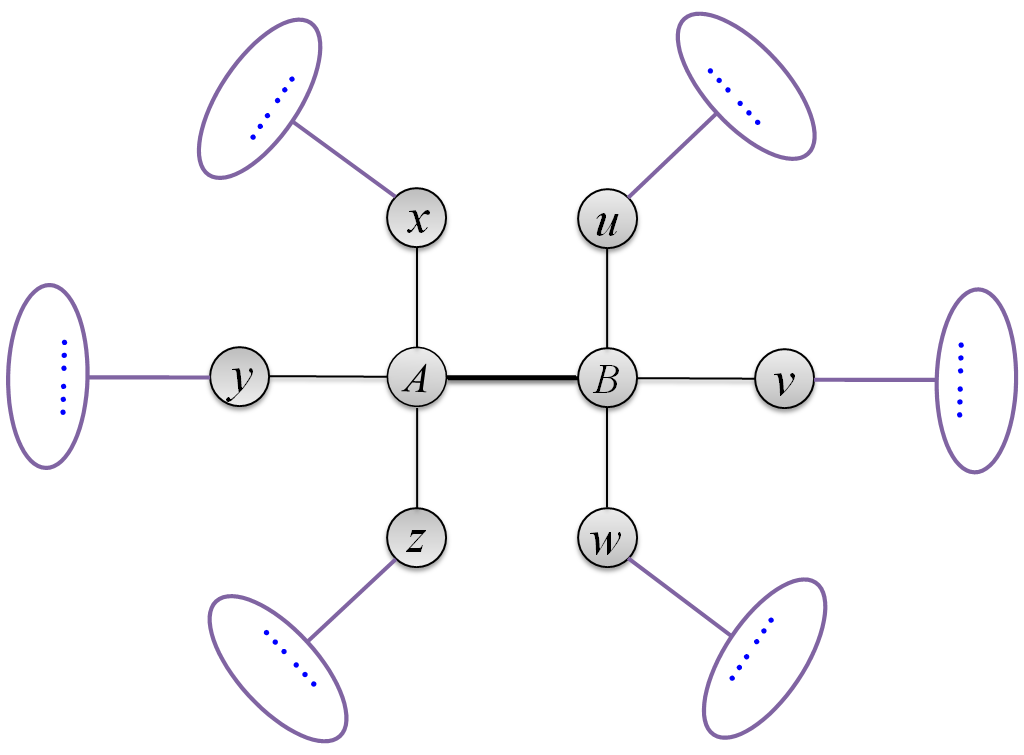}
\caption{(Color online) Schematic illustration of the scheme used for the pair approximation with focal sites $A$ and $B$. This configuration is used to determine changes in the pair configuration probabilities $p_{A,B\rightarrow B,B}$. Since a fraction $\alpha$ of the payoff of each individual reaped from playing games will be shared among the neighborhood, the ultimate gains of an individual are determined by both the strategies of his immediate neighbors and also by those of his next-nearest neighbors. Note that any correlation among those neighbors of $A$ and $B$ (and also their further neighbors) is disregarded by the pair approximation, i.e., all corrections arising from loops (if any) are neglected.\label{fig4}}
\end{figure}

By modifying the dynamical cluster technique~\cite{Hauert2005ajp,Szabo2007pr,Szabo1998pre}, we are able to figure out how the fraction of cooperation evolves as the cost-to-benefit ratio $r$ in the presence of relatedness among the neighboring players. It is worth noting that the (extended) pair approximation is based on the assumption of continuous time, and hence on the asynchronous (i.e., random sequential) strategy-updating~\cite{Hauert2005ajp,Szabo2007pr,Szabo1998pre}. In fact, we have simulated our model with asynchronous strategy-updating, and the extended pair-approximation indeed matches the results better (please see the Supplemental Materials~[\onlinecite{abc}]). For the sake of clarity, figure~\ref{fig4} illustrates a small part of the square lattice with four neighbors. The strategy of the player $A$ is updated by comparing his performance to a randomly chosen neighbor $B$. The payoffs $P_A$ and $P_B$ of $A$ and $B$ are determined by accumulating the payoffs in interactions with their neighbors $x$, $y$, $z$, $B$ and $u$, $v$, $w$, $A$, respectively. The pair approximation is completed by determining the evolution of the pair configuration probabilities, that is, the probability that the pair $p_{A,B}$ becomes $p_{B,B}$:
\begin{widetext}
\begin{equation}
p_{A,B\rightarrow B,B}=\sum_{x,y,z}\sum_{u,v,w}f(P_B-P_A)\times \frac{p_{x,A}p_{y,A}p_{z,A}p_{A,B}p_{u,B}p_{v,B}p_{w,B}}{p_A^3p_B^3}
\end{equation}
\end{widetext}
where the transition probability $f(P_B-P_A)$ is multiplied by the configuration probability and summed over all possible configurations. If $B$ succeeds in populating site $A$, the pair configuration probabilities change: the probabilities $p_{B,B}$, $p_{B,x}$, $p_{B,y}$, and $p_{B,z}$ increase, while the probabilities $p_{A,B}$, $p_{A,x}$, $p_{A,y}$, and $p_{A,z}$ decrease. These changes result in a set of ordinary differential equations:
\begin{widetext}
\begin{eqnarray}
\dot{p}_{c,c}&=&\sum_{xyz}\left[n_c(x,y,z)+1\right]p_{d,x}p_{d,y}p_{d,z}\times \sum_{u,v,w}p_{c,u}p_{c,v}p_{c,w}f\left(P_c(u,v,w,\Omega_u,\Omega_v,\Omega_w) -P_d(x,y,z,\Omega_x,\Omega_y,\Omega_z)\right)-\nonumber\\
&&\sum_{xyz}n_c(x,y,z)p_{c,x}p_{c,y}p_{c,z}\times
\sum_{u,v,w}p_{d,u}p_{d,v}p_{d,w}f\left(P_d(u,v,w,\Omega_u,\Omega_v,\Omega_w) -P_c(x,y,z,\Omega_x,\Omega_y,\Omega_z)\right)\label{a1}\\
\dot{p}_{c,d}&=&\sum_{xyz}\left[(1-n_c(x,y,z))\right]p_{d,x}p_{d,y}p_{d,z}\times \sum_{u,v,w}p_{c,u}p_{c,v}p_{c,w}f\left(P_c(u,v,w,\Omega_u,\Omega_v,\Omega_w) -P_d(x,y,z,\Omega_x,\Omega_y,\Omega_z)\right)-\nonumber\\
&&\sum_{xyz}\left[(2-n_c(x,y,z))\right]p_{c,x}p_{c,y}p_{c,z}\times
\sum_{u,v,w}p_{d,u}p_{d,v}p_{d,w}f\left(P_d(u,v,w,\Omega_u,\Omega_v,\Omega_w) -P_c(x,y,z,\Omega_x,\Omega_y,\Omega_z)\right)\label{a2},
\end{eqnarray}
\end{widetext}
where $n_c(x,y,z)$ is the number of cooperators among the
neighbors $x,y,z$, and $P_c(x,y,z)$ and $P_d(x,y,z)$ specify the
payoffs of a cooperator (defector) interacting with the neighbors
$x,y,z$ plus a defector (cooperator). According the definition of our model in the main text, the payoff function of the individual $A$ depends not only on the strategies of his immediate neighbors, but also on those of his next-nearest neighbors. As an example, here we give out the payoff functions of $A$ and $B$, whom play as a defector and as a cooperator, respectively:
\begin{widetext}
\begin{eqnarray}
P_A&=&P_d(x,y,z,\Omega_x,\Omega_y,\Omega_z)=(1-\alpha)\left[ \left(n_c(x,y,z)+1\right)T+\left(\mathcal{Z}-1-n_c(x,y,z)\right)P \right]+ \nonumber\\
&&\frac{\alpha}{\mathcal{Z}}\left[n_c(x,y,z)\left((\mathcal{Z}-1)(\frac{p_{cd}p_{cc}} {p_c}R+\frac{p_{cd}p_{cd}}{p_c}S)+S\right)+(\mathcal{Z}-1-n_c(x,y,z)) \left((\mathcal{Z}-1)(\frac{p_{dd}p_{cd}}{p_d}T+\frac{p_{dd}p_{dd}}{p_d}P)+P\right) \right]\nonumber\\
P_B&=&P_c(u,v,w,\Omega_u,\Omega_v,\Omega_w)=(1-\alpha)\left[ n_c(u,v,w)R+\left(\mathcal{Z}-1-n_c(u,v,w)\right)S\right]+\nonumber\\
&&\frac{\alpha}{\mathcal{Z}}\left[n_c(u,v,w)\left((\mathcal{Z}-1)(\frac{p_{cc}p_{cc}} {p_c}R+\frac{p_{cc}p_{cd}}{p_c}S)+R\right)+(\mathcal{Z}-1-n_c(u,v,w)) \left((\mathcal{Z}-1)(\frac{p_{cd}p_{cd}}{p_d}T+\frac{p_{cd}p_{dd}}{p_d}P)+T\right) \right]\nonumber
\end{eqnarray}
\end{widetext}
For simplicity, we only consider the case where each individual interacts with other $\mathcal{Z}$ individuals and their common neighbors (if any) are considered to be independent, i.e., we consider a Cayley tree with coordination number $\mathcal{Z}$. The above two equations, Eqs.~(\ref{a1}) and (\ref{a2}), omit the common factor $2p_{c,d}/(p_c^3p_d^3)$ which can be absorbed into the time measurement units. In combination with the symmetry condition $p_{c,d}=p_{d,c}$ and the constraint $p_{c,c}+p_{c,d}+p_{d,c}+p_{d,d}=1$, these two equations can be treated either by numerical integration or by setting $\dot p_{c,c}=\dot p_{c,d}=0$ and solving for $p_{c,c}$ and $p_{c,d}$. Then the equilibrium fraction of cooperators in the whole population is obtained by $p_c=p_{c,c}+p_{c,d}$.

\bibliography{ref}

\begin{thebibliography}{66}%
\makeatletter
\providecommand \@ifxundefined [1]{%
 \@ifx{#1\undefined}
}%
\providecommand \@ifnum [1]{%
 \ifnum #1\expandafter \@firstoftwo
 \else \expandafter \@secondoftwo
 \fi
}%
\providecommand \@ifx [1]{%
 \ifx #1\expandafter \@firstoftwo
 \else \expandafter \@secondoftwo
 \fi
}%
\providecommand \natexlab [1]{#1}%
\providecommand \enquote  [1]{``#1''}%
\providecommand \bibnamefont  [1]{#1}%
\providecommand \bibfnamefont [1]{#1}%
\providecommand \citenamefont [1]{#1}%
\providecommand \href@noop [0]{\@secondoftwo}%
\providecommand \href [0]{\begingroup \@sanitize@url \@href}%
\providecommand \@href[1]{\@@startlink{#1}\@@href}%
\providecommand \@@href[1]{\endgroup#1\@@endlink}%
\providecommand \@sanitize@url [0]{\catcode `\\12\catcode `\$12\catcode
  `\&12\catcode `\#12\catcode `\^12\catcode `\_12\catcode `\%12\relax}%
\providecommand \@@startlink[1]{}%
\providecommand \@@endlink[0]{}%
\providecommand \url  [0]{\begingroup\@sanitize@url \@url }%
\providecommand \@url [1]{\endgroup\@href {#1}{\urlprefix }}%
\providecommand \urlprefix  [0]{URL }%
\providecommand \Eprint [0]{\href }%
\providecommand \doibase [0]{http://dx.doi.org/}%
\providecommand \selectlanguage [0]{\@gobble}%
\providecommand \bibinfo  [0]{\@secondoftwo}%
\providecommand \bibfield  [0]{\@secondoftwo}%
\providecommand \translation [1]{[#1]}%
\providecommand \BibitemOpen [0]{}%
\providecommand \bibitemStop [0]{}%
\providecommand \bibitemNoStop [0]{.\EOS\space}%
\providecommand \EOS [0]{\spacefactor3000\relax}%
\providecommand \BibitemShut  [1]{\csname bibitem#1\endcsname}%
\let\auto@bib@innerbib\@empty
\bibitem [{\citenamefont {Axelrod}(2006)}]{Axelrod2006book}%
  \BibitemOpen
  \bibfield  {author} {\bibinfo {author} {\bibfnamefont {R.~M.}\ \bibnamefont
  {Axelrod}},\ }\href@noop {} {\emph {\bibinfo {title} {The Evolution of
  Cooperation}}}\ (\bibinfo  {publisher} {Perseus Books, New York},\ \bibinfo
  {year} {2006})\ \bibinfo {note} {revised ed.}\BibitemShut {Stop}%
\bibitem [{\citenamefont {Nowak}(2006{\natexlab{a}})}]{Nowak2006science}%
  \BibitemOpen
  \bibfield  {author} {\bibinfo {author} {\bibfnamefont {M.~A.}\ \bibnamefont
  {Nowak}},\ }\href {\doibase 10.1126/science.1133755} {\bibfield  {journal}
  {\bibinfo  {journal} {Science}\ }\textbf {\bibinfo {volume} {314}},\ \bibinfo
  {pages} {1560} (\bibinfo {year} {2006}{\natexlab{a}})}\BibitemShut {NoStop}%
\bibitem [{\citenamefont {Maynard~Smith}\ and\ \citenamefont
  {Price}(1973)}]{Smith1973nature}%
  \BibitemOpen
  \bibfield  {author} {\bibinfo {author} {\bibfnamefont {J.}~\bibnamefont
  {Maynard~Smith}}\ and\ \bibinfo {author} {\bibfnamefont {G.~R.}\ \bibnamefont
  {Price}},\ }\href {\doibase 10.1038/246015a0} {\bibfield  {journal} {\bibinfo
   {journal} {Nature}\ }\textbf {\bibinfo {volume} {246}},\ \bibinfo {pages}
  {15} (\bibinfo {year} {1973})}\BibitemShut {NoStop}%
\bibitem [{\citenamefont {Axelrod}\ and\ \citenamefont
  {Hamilton}(1981)}]{Axelrod1981science}%
  \BibitemOpen
  \bibfield  {author} {\bibinfo {author} {\bibfnamefont {R.}~\bibnamefont
  {Axelrod}}\ and\ \bibinfo {author} {\bibfnamefont {W.~D.}\ \bibnamefont
  {Hamilton}},\ }\href {\doibase 10.1126/science.7466396} {\bibfield  {journal}
  {\bibinfo  {journal} {Science}\ }\textbf {\bibinfo {volume} {211}},\ \bibinfo
  {pages} {1390} (\bibinfo {year} {1981})}\BibitemShut {NoStop}%
\bibitem [{\citenamefont {Hofbauer}\ and\ \citenamefont
  {Sigmund}(1998)}]{Hofbauer1998book}%
  \BibitemOpen
  \bibfield  {author} {\bibinfo {author} {\bibfnamefont {J.}~\bibnamefont
  {Hofbauer}}\ and\ \bibinfo {author} {\bibfnamefont {K.}~\bibnamefont
  {Sigmund}},\ }\href@noop {} {\emph {\bibinfo {title} {Evolutionary Games and
  Population Dynamics}}}\ (\bibinfo  {publisher} {Cambridge University Press,
  Cambridge, UK},\ \bibinfo {year} {1998})\BibitemShut {NoStop}%
\bibitem [{\citenamefont {Nowak}(2006{\natexlab{b}})}]{Nowak2006book}%
  \BibitemOpen
  \bibfield  {author} {\bibinfo {author} {\bibfnamefont {M.}~\bibnamefont
  {Nowak}},\ }\href@noop {} {\emph {\bibinfo {title} {Evolutionary Dynamics:
  Exploring the Equations of Life}}}\ (\bibinfo  {publisher} {Harvard
  University Press, Cambridge, MA},\ \bibinfo {year} {2006})\BibitemShut
  {NoStop}%
\bibitem [{\citenamefont {Sigmund}(2010)}]{Sigmund2010book}%
  \BibitemOpen
  \bibfield  {author} {\bibinfo {author} {\bibfnamefont {K.}~\bibnamefont
  {Sigmund}},\ }\href@noop {} {\emph {\bibinfo {title} {The Calculus of
  Selfishness:}}},\ Princeton Series in Theoretical and Computational Biology\
  (\bibinfo  {publisher} {Princeton University Press, Princeton, NJ},\ \bibinfo
  {year} {2010})\BibitemShut {NoStop}%
\bibitem [{\citenamefont {Maynard~Smith}(1982)}]{Maynard1982book}%
  \BibitemOpen
  \bibfield  {author} {\bibinfo {author} {\bibfnamefont {J.}~\bibnamefont
  {Maynard~Smith}},\ }\href@noop {} {\emph {\bibinfo {title} {Evolution and the
  Theory of Games}}}\ (\bibinfo  {publisher} {Cambridge University Press,
  Cambridge},\ \bibinfo {year} {1982})\BibitemShut {NoStop}%
\bibitem [{\citenamefont {K{\"u}mmerli}\ \emph {et~al.}(2007)\citenamefont
  {K{\"u}mmerli}, \citenamefont {Colliard}, \citenamefont {Fiechter},
  \citenamefont {Petitpierre}, \citenamefont {Russier},\ and\ \citenamefont
  {Keller}}]{Kummerli2007prsb}%
  \BibitemOpen
  \bibfield  {author} {\bibinfo {author} {\bibfnamefont {R.}~\bibnamefont
  {K{\"u}mmerli}}, \bibinfo {author} {\bibfnamefont {C.}~\bibnamefont
  {Colliard}}, \bibinfo {author} {\bibfnamefont {N.}~\bibnamefont {Fiechter}},
  \bibinfo {author} {\bibfnamefont {B.}~\bibnamefont {Petitpierre}}, \bibinfo
  {author} {\bibfnamefont {F.}~\bibnamefont {Russier}}, \ and\ \bibinfo
  {author} {\bibfnamefont {L.}~\bibnamefont {Keller}},\ }\href {\doibase
  10.1098/rspb.2007.0793} {\bibfield  {journal} {\bibinfo  {journal}
  {Proceedings of the Royal Society B: Biological Sciences}\ }\textbf {\bibinfo
  {volume} {274}},\ \bibinfo {pages} {2965} (\bibinfo {year}
  {2007})}\BibitemShut {NoStop}%
\bibitem [{\citenamefont {Hauert}\ and\ \citenamefont
  {Doebeli}(2004)}]{Hauert2004nature}%
  \BibitemOpen
  \bibfield  {author} {\bibinfo {author} {\bibfnamefont {C.}~\bibnamefont
  {Hauert}}\ and\ \bibinfo {author} {\bibfnamefont {M.}~\bibnamefont
  {Doebeli}},\ }\href {\doibase 10.1038/nature02360} {\bibfield  {journal}
  {\bibinfo  {journal} {Nature}\ }\textbf {\bibinfo {volume} {428}},\ \bibinfo
  {pages} {643} (\bibinfo {year} {2004})}\BibitemShut {NoStop}%
\bibitem [{\citenamefont {Hamilton}(1964{\natexlab{a}})}]{Hamilton1964jtb1}%
  \BibitemOpen
  \bibfield  {author} {\bibinfo {author} {\bibfnamefont {W.}~\bibnamefont
  {Hamilton}},\ }\href {\doibase
  http://dx.doi.org/10.1016/0022-5193(64)90039-6} {\bibfield  {journal}
  {\bibinfo  {journal} {Journal of Theoretical Biology}\ }\textbf {\bibinfo
  {volume} {7}},\ \bibinfo {pages} {17} (\bibinfo {year}
  {1964}{\natexlab{a}})}\BibitemShut {NoStop}%
\bibitem [{\citenamefont {Hamilton}(1964{\natexlab{b}})}]{Hamilton1964jtb}%
  \BibitemOpen
  \bibfield  {author} {\bibinfo {author} {\bibfnamefont {W.}~\bibnamefont
  {Hamilton}},\ }\href {\doibase
  http://dx.doi.org/10.1016/0022-5193(64)90038-4} {\bibfield  {journal}
  {\bibinfo  {journal} {Journal of Theoretical Biology}\ }\textbf {\bibinfo
  {volume} {7}},\ \bibinfo {pages} {1 } (\bibinfo {year}
  {1964}{\natexlab{b}})}\BibitemShut {NoStop}%
\bibitem [{\citenamefont {Trivers}(1971)}]{Robert1971qrb}%
  \BibitemOpen
  \bibfield  {author} {\bibinfo {author} {\bibfnamefont {R.~L.}\ \bibnamefont
  {Trivers}},\ }\href {http://www.jstor.org/stable/2822435} {\bibfield
  {journal} {\bibinfo  {journal} {The Quarterly Review of Biology}\ }\textbf
  {\bibinfo {volume} {46}},\ \bibinfo {pages} {35} (\bibinfo {year}
  {1971})}\BibitemShut {NoStop}%
\bibitem [{\citenamefont {Nowak}\ and\ \citenamefont
  {Sigmund}(2005)}]{Nowak2005nature}%
  \BibitemOpen
  \bibfield  {author} {\bibinfo {author} {\bibfnamefont {M.~A.}\ \bibnamefont
  {Nowak}}\ and\ \bibinfo {author} {\bibfnamefont {K.}~\bibnamefont
  {Sigmund}},\ }\href {\doibase 10.1038/nature04131} {\bibfield  {journal}
  {\bibinfo  {journal} {Nature}\ }\textbf {\bibinfo {volume} {437}},\ \bibinfo
  {pages} {1291} (\bibinfo {year} {2005})}\BibitemShut {NoStop}%
\bibitem [{\citenamefont {Nowak}\ and\ \citenamefont
  {May}(1992)}]{Nowak1992nature}%
  \BibitemOpen
  \bibfield  {author} {\bibinfo {author} {\bibfnamefont {M.~A.}\ \bibnamefont
  {Nowak}}\ and\ \bibinfo {author} {\bibfnamefont {R.~M.}\ \bibnamefont
  {May}},\ }\href {\doibase 10.1038/359826a0} {\bibfield  {journal} {\bibinfo
  {journal} {Nature}\ }\textbf {\bibinfo {volume} {359}},\ \bibinfo {pages}
  {826} (\bibinfo {year} {1992})}\BibitemShut {NoStop}%
\bibitem [{\citenamefont {Szab\'o}\ and\ \citenamefont
  {T\ifmmode~\mbox{\H{o}}\else \H{o}\fi{}ke}(1998)}]{Szabo1998pre}%
  \BibitemOpen
  \bibfield  {author} {\bibinfo {author} {\bibfnamefont {G.}~\bibnamefont
  {Szab\'o}}\ and\ \bibinfo {author} {\bibfnamefont {C.}~\bibnamefont
  {T\ifmmode~\mbox{\H{o}}\else \H{o}\fi{}ke}},\ }\href {\doibase
  10.1103/PhysRevE.58.69} {\bibfield  {journal} {\bibinfo  {journal} {Phys.
  Rev. E}\ }\textbf {\bibinfo {volume} {58}},\ \bibinfo {pages} {69} (\bibinfo
  {year} {1998})}\BibitemShut {NoStop}%
\bibitem [{\citenamefont {Ohtsuki}\ \emph {et~al.}(2006)\citenamefont
  {Ohtsuki}, \citenamefont {Hauert},\ and\ \citenamefont
  {Nowak}}]{Ohtsuki2006nature}%
  \BibitemOpen
  \bibfield  {author} {\bibinfo {author} {\bibfnamefont {H.}~\bibnamefont
  {Ohtsuki}}, \bibinfo {author} {\bibfnamefont {C.}~\bibnamefont {Hauert}}, \
  and\ \bibinfo {author} {\bibfnamefont {M.~A.}\ \bibnamefont {Nowak}},\ }\href
  {\doibase 10.1038/nature04605} {\bibfield  {journal} {\bibinfo  {journal}
  {Nature}\ }\textbf {\bibinfo {volume} {441}},\ \bibinfo {pages} {502}
  (\bibinfo {year} {2006})}\BibitemShut {NoStop}%
\bibitem [{\citenamefont {Wilson}(1975)}]{Wilson1975pnas}%
  \BibitemOpen
  \bibfield  {author} {\bibinfo {author} {\bibfnamefont {D.~S.}\ \bibnamefont
  {Wilson}},\ }\href {http://www.pnas.org/content/72/1/143.abstract} {\bibfield
   {journal} {\bibinfo  {journal} {Proceedings of the National Academy of
  Sciences}\ }\textbf {\bibinfo {volume} {72}},\ \bibinfo {pages} {143}
  (\bibinfo {year} {1975})}\BibitemShut {NoStop}%
\bibitem [{\citenamefont {Traulsen}\ and\ \citenamefont
  {Nowak}(2006)}]{Traulsen2006pnas}%
  \BibitemOpen
  \bibfield  {author} {\bibinfo {author} {\bibfnamefont {A.}~\bibnamefont
  {Traulsen}}\ and\ \bibinfo {author} {\bibfnamefont {M.~A.}\ \bibnamefont
  {Nowak}},\ }\href {\doibase 10.1073/pnas.0602530103} {\bibfield  {journal}
  {\bibinfo  {journal} {Proceedings of the National Academy of Sciences}\
  }\textbf {\bibinfo {volume} {103}},\ \bibinfo {pages} {10952} (\bibinfo
  {year} {2006})}\BibitemShut {NoStop}%
\bibitem [{\citenamefont {Szab\'o}\ and\ \citenamefont
  {Hauert}(2002)}]{Szabo2002prl}%
  \BibitemOpen
  \bibfield  {author} {\bibinfo {author} {\bibfnamefont {G.}~\bibnamefont
  {Szab\'o}}\ and\ \bibinfo {author} {\bibfnamefont {C.}~\bibnamefont
  {Hauert}},\ }\href {\doibase 10.1103/PhysRevLett.89.118101} {\bibfield
  {journal} {\bibinfo  {journal} {Phys. Rev. Lett.}\ }\textbf {\bibinfo
  {volume} {89}},\ \bibinfo {pages} {118101} (\bibinfo {year}
  {2002})}\BibitemShut {NoStop}%
\bibitem [{\citenamefont {Wu}\ \emph {et~al.}(2006)\citenamefont {Wu},
  \citenamefont {Xu}, \citenamefont {Huang}, \citenamefont {Wang},\ and\
  \citenamefont {Wang}}]{Wu2006pre}%
  \BibitemOpen
  \bibfield  {author} {\bibinfo {author} {\bibfnamefont {Z.-X.}\ \bibnamefont
  {Wu}}, \bibinfo {author} {\bibfnamefont {X.-J.}\ \bibnamefont {Xu}}, \bibinfo
  {author} {\bibfnamefont {Z.-G.}\ \bibnamefont {Huang}}, \bibinfo {author}
  {\bibfnamefont {S.-J.}\ \bibnamefont {Wang}}, \ and\ \bibinfo {author}
  {\bibfnamefont {Y.-H.}\ \bibnamefont {Wang}},\ }\href {\doibase
  10.1103/PhysRevE.74.021107} {\bibfield  {journal} {\bibinfo  {journal} {Phys.
  Rev. E}\ }\textbf {\bibinfo {volume} {74}},\ \bibinfo {pages} {021107}
  (\bibinfo {year} {2006})}\BibitemShut {NoStop}%
\bibitem [{\citenamefont {Santos}\ and\ \citenamefont
  {Pacheco}(2005)}]{Santos2005prl}%
  \BibitemOpen
  \bibfield  {author} {\bibinfo {author} {\bibfnamefont {F.~C.}\ \bibnamefont
  {Santos}}\ and\ \bibinfo {author} {\bibfnamefont {J.~M.}\ \bibnamefont
  {Pacheco}},\ }\href {\doibase 10.1103/PhysRevLett.95.098104} {\bibfield
  {journal} {\bibinfo  {journal} {Phys. Rev. Lett.}\ }\textbf {\bibinfo
  {volume} {95}},\ \bibinfo {pages} {098104} (\bibinfo {year}
  {2005})}\BibitemShut {NoStop}%
\bibitem [{\citenamefont {G\'omez-Garde\~nes}\ \emph
  {et~al.}(2007)\citenamefont {G\'omez-Garde\~nes}, \citenamefont {Campillo},
  \citenamefont {Flor\'ia},\ and\ \citenamefont {Moreno}}]{Gardenes2007prl}%
  \BibitemOpen
  \bibfield  {author} {\bibinfo {author} {\bibfnamefont {J.}~\bibnamefont
  {G\'omez-Garde\~nes}}, \bibinfo {author} {\bibfnamefont {M.}~\bibnamefont
  {Campillo}}, \bibinfo {author} {\bibfnamefont {L.~M.}\ \bibnamefont
  {Flor\'ia}}, \ and\ \bibinfo {author} {\bibfnamefont {Y.}~\bibnamefont
  {Moreno}},\ }\href {\doibase 10.1103/PhysRevLett.98.108103} {\bibfield
  {journal} {\bibinfo  {journal} {Phys. Rev. Lett.}\ }\textbf {\bibinfo
  {volume} {98}},\ \bibinfo {pages} {108103} (\bibinfo {year}
  {2007})}\BibitemShut {NoStop}%
\bibitem [{\citenamefont {Rong}\ \emph {et~al.}(2007)\citenamefont {Rong},
  \citenamefont {Li},\ and\ \citenamefont {Wang}}]{Rong2007pre}%
  \BibitemOpen
  \bibfield  {author} {\bibinfo {author} {\bibfnamefont {Z.}~\bibnamefont
  {Rong}}, \bibinfo {author} {\bibfnamefont {X.}~\bibnamefont {Li}}, \ and\
  \bibinfo {author} {\bibfnamefont {X.}~\bibnamefont {Wang}},\ }\href {\doibase
  10.1103/PhysRevE.76.027101} {\bibfield  {journal} {\bibinfo  {journal} {Phys.
  Rev. E}\ }\textbf {\bibinfo {volume} {76}},\ \bibinfo {pages} {027101}
  (\bibinfo {year} {2007})}\BibitemShut {NoStop}%
\bibitem [{\citenamefont {Szolnoki}\ and\ \citenamefont
  {Szab\'o}(2007)}]{Szolnoki2007epl}%
  \BibitemOpen
  \bibfield  {author} {\bibinfo {author} {\bibfnamefont {A.}~\bibnamefont
  {Szolnoki}}\ and\ \bibinfo {author} {\bibfnamefont {G.}~\bibnamefont
  {Szab\'o}},\ }\href {\doibase 10.1209/0295-5075/77/30004} {\bibfield
  {journal} {\bibinfo  {journal} {EPL}\ }\textbf {\bibinfo {volume} {77}},\
  \bibinfo {pages} {30004} (\bibinfo {year} {2007})}\BibitemShut {NoStop}%
\bibitem [{\citenamefont {Szab\'o}\ and\ \citenamefont
  {Szolnoki}(2009)}]{Szabo2009pre}%
  \BibitemOpen
  \bibfield  {author} {\bibinfo {author} {\bibfnamefont {G.}~\bibnamefont
  {Szab\'o}}\ and\ \bibinfo {author} {\bibfnamefont {A.}~\bibnamefont
  {Szolnoki}},\ }\href {\doibase 10.1103/PhysRevE.79.016106} {\bibfield
  {journal} {\bibinfo  {journal} {Phys. Rev. E}\ }\textbf {\bibinfo {volume}
  {79}},\ \bibinfo {pages} {016106} (\bibinfo {year} {2009})}\BibitemShut
  {NoStop}%
\bibitem [{\citenamefont {Pacheco}\ \emph {et~al.}(2006)\citenamefont
  {Pacheco}, \citenamefont {Traulsen},\ and\ \citenamefont
  {Nowak}}]{Pacheco2006prl}%
  \BibitemOpen
  \bibfield  {author} {\bibinfo {author} {\bibfnamefont {J.~M.}\ \bibnamefont
  {Pacheco}}, \bibinfo {author} {\bibfnamefont {A.}~\bibnamefont {Traulsen}}, \
  and\ \bibinfo {author} {\bibfnamefont {M.~A.}\ \bibnamefont {Nowak}},\ }\href
  {\doibase 10.1103/PhysRevLett.97.258103} {\bibfield  {journal} {\bibinfo
  {journal} {Phys. Rev. Lett.}\ }\textbf {\bibinfo {volume} {97}},\ \bibinfo
  {pages} {258103} (\bibinfo {year} {2006})}\BibitemShut {NoStop}%
\bibitem [{\citenamefont {Ohtsuki}\ \emph {et~al.}(2007)\citenamefont
  {Ohtsuki}, \citenamefont {Nowak},\ and\ \citenamefont
  {Pacheco}}]{Ohtsuki2007prl}%
  \BibitemOpen
  \bibfield  {author} {\bibinfo {author} {\bibfnamefont {H.}~\bibnamefont
  {Ohtsuki}}, \bibinfo {author} {\bibfnamefont {M.~A.}\ \bibnamefont {Nowak}},
  \ and\ \bibinfo {author} {\bibfnamefont {J.~M.}\ \bibnamefont {Pacheco}},\
  }\href {\doibase 10.1103/PhysRevLett.98.108106} {\bibfield  {journal}
  {\bibinfo  {journal} {Phys. Rev. Lett.}\ }\textbf {\bibinfo {volume} {98}},\
  \bibinfo {pages} {108106} (\bibinfo {year} {2007})}\BibitemShut {NoStop}%
\bibitem [{\citenamefont {Wu}\ and\ \citenamefont {Wang}(2007)}]{Wu2007pre}%
  \BibitemOpen
  \bibfield  {author} {\bibinfo {author} {\bibfnamefont {Z.-X.}\ \bibnamefont
  {Wu}}\ and\ \bibinfo {author} {\bibfnamefont {Y.-H.}\ \bibnamefont {Wang}},\
  }\href {\doibase 10.1103/PhysRevE.75.041114} {\bibfield  {journal} {\bibinfo
  {journal} {Phys. Rev. E}\ }\textbf {\bibinfo {volume} {75}},\ \bibinfo
  {pages} {041114} (\bibinfo {year} {2007})}\BibitemShut {NoStop}%
\bibitem [{\citenamefont {Roca}\ \emph {et~al.}(2006)\citenamefont {Roca},
  \citenamefont {Cuesta},\ and\ \citenamefont {S\'anchez}}]{Roca2006prl}%
  \BibitemOpen
  \bibfield  {author} {\bibinfo {author} {\bibfnamefont {C.~P.}\ \bibnamefont
  {Roca}}, \bibinfo {author} {\bibfnamefont {J.~A.}\ \bibnamefont {Cuesta}}, \
  and\ \bibinfo {author} {\bibfnamefont {A.}~\bibnamefont {S\'anchez}},\ }\href
  {\doibase 10.1103/PhysRevLett.97.158701} {\bibfield  {journal} {\bibinfo
  {journal} {Phys. Rev. Lett.}\ }\textbf {\bibinfo {volume} {97}},\ \bibinfo
  {pages} {158701} (\bibinfo {year} {2006})}\BibitemShut {NoStop}%
\bibitem [{\citenamefont {Wu}\ \emph {et~al.}(2009)\citenamefont {Wu},
  \citenamefont {Rong},\ and\ \citenamefont {Holme}}]{Wu2009pre}%
  \BibitemOpen
  \bibfield  {author} {\bibinfo {author} {\bibfnamefont {Z.-X.}\ \bibnamefont
  {Wu}}, \bibinfo {author} {\bibfnamefont {Z.}~\bibnamefont {Rong}}, \ and\
  \bibinfo {author} {\bibfnamefont {P.}~\bibnamefont {Holme}},\ }\href
  {\doibase 10.1103/PhysRevE.80.036106} {\bibfield  {journal} {\bibinfo
  {journal} {Phys. Rev. E}\ }\textbf {\bibinfo {volume} {80}},\ \bibinfo
  {pages} {036106} (\bibinfo {year} {2009})}\BibitemShut {NoStop}%
\bibitem [{\citenamefont {Rong}\ \emph {et~al.}(2010)\citenamefont {Rong},
  \citenamefont {Wu},\ and\ \citenamefont {Wang}}]{Rong2010pre}%
  \BibitemOpen
  \bibfield  {author} {\bibinfo {author} {\bibfnamefont {Z.}~\bibnamefont
  {Rong}}, \bibinfo {author} {\bibfnamefont {Z.-X.}\ \bibnamefont {Wu}}, \ and\
  \bibinfo {author} {\bibfnamefont {W.-X.}\ \bibnamefont {Wang}},\ }\href
  {\doibase 10.1103/PhysRevE.82.026101} {\bibfield  {journal} {\bibinfo
  {journal} {Phys. Rev. E}\ }\textbf {\bibinfo {volume} {82}},\ \bibinfo
  {pages} {026101} (\bibinfo {year} {2010})}\BibitemShut {NoStop}%
\bibitem [{\citenamefont {Chen}\ and\ \citenamefont
  {Wang}(2008)}]{Chen2008pre}%
  \BibitemOpen
  \bibfield  {author} {\bibinfo {author} {\bibfnamefont {X.}~\bibnamefont
  {Chen}}\ and\ \bibinfo {author} {\bibfnamefont {L.}~\bibnamefont {Wang}},\
  }\href {\doibase 10.1103/PhysRevE.77.017103} {\bibfield  {journal} {\bibinfo
  {journal} {Phys. Rev. E}\ }\textbf {\bibinfo {volume} {77}},\ \bibinfo
  {pages} {017103} (\bibinfo {year} {2008})}\BibitemShut {NoStop}%
\bibitem [{\citenamefont {Perc}\ and\ \citenamefont
  {Szolnoki}(2008)}]{Perc2008pre}%
  \BibitemOpen
  \bibfield  {author} {\bibinfo {author} {\bibfnamefont {M.}~\bibnamefont
  {Perc}}\ and\ \bibinfo {author} {\bibfnamefont {A.}~\bibnamefont
  {Szolnoki}},\ }\href {\doibase 10.1103/PhysRevE.77.011904} {\bibfield
  {journal} {\bibinfo  {journal} {Phys. Rev. E}\ }\textbf {\bibinfo {volume}
  {77}},\ \bibinfo {pages} {011904} (\bibinfo {year} {2008})}\BibitemShut
  {NoStop}%
\bibitem [{\citenamefont {Yang}\ \emph {et~al.}(2009)\citenamefont {Yang},
  \citenamefont {Wang}, \citenamefont {Wu}, \citenamefont {Lai},\ and\
  \citenamefont {Wang}}]{Yang2009pre}%
  \BibitemOpen
  \bibfield  {author} {\bibinfo {author} {\bibfnamefont {H.-X.}\ \bibnamefont
  {Yang}}, \bibinfo {author} {\bibfnamefont {W.-X.}\ \bibnamefont {Wang}},
  \bibinfo {author} {\bibfnamefont {Z.-X.}\ \bibnamefont {Wu}}, \bibinfo
  {author} {\bibfnamefont {Y.-C.}\ \bibnamefont {Lai}}, \ and\ \bibinfo
  {author} {\bibfnamefont {B.-H.}\ \bibnamefont {Wang}},\ }\href {\doibase
  10.1103/PhysRevE.79.056107} {\bibfield  {journal} {\bibinfo  {journal} {Phys.
  Rev. E}\ }\textbf {\bibinfo {volume} {79}},\ \bibinfo {pages} {056107}
  (\bibinfo {year} {2009})}\BibitemShut {NoStop}%
\bibitem [{\citenamefont {Yang}\ \emph {et~al.}(2010)\citenamefont {Yang},
  \citenamefont {Wu},\ and\ \citenamefont {Wang}}]{Yang2010pre}%
  \BibitemOpen
  \bibfield  {author} {\bibinfo {author} {\bibfnamefont {H.-X.}\ \bibnamefont
  {Yang}}, \bibinfo {author} {\bibfnamefont {Z.-X.}\ \bibnamefont {Wu}}, \ and\
  \bibinfo {author} {\bibfnamefont {B.-H.}\ \bibnamefont {Wang}},\ }\href
  {\doibase 10.1103/PhysRevE.81.065101} {\bibfield  {journal} {\bibinfo
  {journal} {Phys. Rev. E}\ }\textbf {\bibinfo {volume} {81}},\ \bibinfo
  {pages} {065101} (\bibinfo {year} {2010})}\BibitemShut {NoStop}%
\bibitem [{\citenamefont {Jiang}\ \emph {et~al.}(2010)\citenamefont {Jiang},
  \citenamefont {Wang}, \citenamefont {Lai},\ and\ \citenamefont
  {Wang}}]{Jiang2010pre}%
  \BibitemOpen
  \bibfield  {author} {\bibinfo {author} {\bibfnamefont {L.-L.}\ \bibnamefont
  {Jiang}}, \bibinfo {author} {\bibfnamefont {W.-X.}\ \bibnamefont {Wang}},
  \bibinfo {author} {\bibfnamefont {Y.-C.}\ \bibnamefont {Lai}}, \ and\
  \bibinfo {author} {\bibfnamefont {B.-H.}\ \bibnamefont {Wang}},\ }\href
  {\doibase 10.1103/PhysRevE.81.036108} {\bibfield  {journal} {\bibinfo
  {journal} {Phys. Rev. E}\ }\textbf {\bibinfo {volume} {81}},\ \bibinfo
  {pages} {036108} (\bibinfo {year} {2010})}\BibitemShut {NoStop}%
\bibitem [{\citenamefont {Roca}\ and\ \citenamefont
  {Helbing}(2011)}]{Roca2011pnas}%
  \BibitemOpen
  \bibfield  {author} {\bibinfo {author} {\bibfnamefont {C.~P.}\ \bibnamefont
  {Roca}}\ and\ \bibinfo {author} {\bibfnamefont {D.}~\bibnamefont {Helbing}},\
  }\href {\doibase 10.1073/pnas.1101044108} {\bibfield  {journal} {\bibinfo
  {journal} {Proceedings of the National Academy of Sciences}\ }\textbf
  {\bibinfo {volume} {108}},\ \bibinfo {pages} {11370} (\bibinfo {year}
  {2011})}\BibitemShut {NoStop}%
\bibitem [{\citenamefont {Chen}\ \emph {et~al.}(2012)\citenamefont {Chen},
  \citenamefont {Szolnoki},\ and\ \citenamefont {Perc}}]{Chen2012pre}%
  \BibitemOpen
  \bibfield  {author} {\bibinfo {author} {\bibfnamefont {X.}~\bibnamefont
  {Chen}}, \bibinfo {author} {\bibfnamefont {A.}~\bibnamefont {Szolnoki}}, \
  and\ \bibinfo {author} {\bibfnamefont {M.}~\bibnamefont {Perc}},\ }\href
  {\doibase 10.1103/PhysRevE.86.036101} {\bibfield  {journal} {\bibinfo
  {journal} {Phys. Rev. E}\ }\textbf {\bibinfo {volume} {86}},\ \bibinfo
  {pages} {036101} (\bibinfo {year} {2012})}\BibitemShut {NoStop}%
\bibitem [{\citenamefont {Requejo}\ and\ \citenamefont
  {Camacho}(2012)}]{Requejo2012prl}%
  \BibitemOpen
  \bibfield  {author} {\bibinfo {author} {\bibfnamefont {R.~J.}\ \bibnamefont
  {Requejo}}\ and\ \bibinfo {author} {\bibfnamefont {J.}~\bibnamefont
  {Camacho}},\ }\href {\doibase 10.1103/PhysRevLett.108.038701} {\bibfield
  {journal} {\bibinfo  {journal} {Phys. Rev. Lett.}\ }\textbf {\bibinfo
  {volume} {108}},\ \bibinfo {pages} {038701} (\bibinfo {year}
  {2012})}\BibitemShut {NoStop}%
\bibitem [{\citenamefont {Requejo}\ and\ \citenamefont
  {Camacho}(2013)}]{Requejo2013pre}%
  \BibitemOpen
  \bibfield  {author} {\bibinfo {author} {\bibfnamefont {R.~J.}\ \bibnamefont
  {Requejo}}\ and\ \bibinfo {author} {\bibfnamefont {J.}~\bibnamefont
  {Camacho}},\ }\href {\doibase 10.1103/PhysRevE.87.022819} {\bibfield
  {journal} {\bibinfo  {journal} {Phys. Rev. E}\ }\textbf {\bibinfo {volume}
  {87}},\ \bibinfo {pages} {022819} (\bibinfo {year} {2013})}\BibitemShut
  {NoStop}%
\bibitem [{\citenamefont {Szolnoki}\ and\ \citenamefont
  {Perc}(2012)}]{Szolnoki2012pre}%
  \BibitemOpen
  \bibfield  {author} {\bibinfo {author} {\bibfnamefont {A.}~\bibnamefont
  {Szolnoki}}\ and\ \bibinfo {author} {\bibfnamefont {M.}~\bibnamefont
  {Perc}},\ }\href {\doibase 10.1103/PhysRevE.85.026104} {\bibfield  {journal}
  {\bibinfo  {journal} {Phys. Rev. E}\ }\textbf {\bibinfo {volume} {85}},\
  \bibinfo {pages} {026104} (\bibinfo {year} {2012})}\BibitemShut {NoStop}%
\bibitem [{\citenamefont {Wang}\ \emph {et~al.}(2011)\citenamefont {Wang},
  \citenamefont {Murks}, \citenamefont {Du}, \citenamefont {Rong},\ and\
  \citenamefont {Perc}}]{Wang2011jtb}%
  \BibitemOpen
  \bibfield  {author} {\bibinfo {author} {\bibfnamefont {Z.}~\bibnamefont
  {Wang}}, \bibinfo {author} {\bibfnamefont {A.}~\bibnamefont {Murks}},
  \bibinfo {author} {\bibfnamefont {W.-B.}\ \bibnamefont {Du}}, \bibinfo
  {author} {\bibfnamefont {Z.-H.}\ \bibnamefont {Rong}}, \ and\ \bibinfo
  {author} {\bibfnamefont {M.}~\bibnamefont {Perc}},\ }\href {\doibase
  http://dx.doi.org/10.1016/j.jtbi.2011.02.016} {\bibfield  {journal} {\bibinfo
   {journal} {Journal of Theoretical Biology}\ }\textbf {\bibinfo {volume}
  {277}},\ \bibinfo {pages} {19 } (\bibinfo {year} {2011})}\BibitemShut
  {NoStop}%
\bibitem [{\citenamefont {Rand}\ \emph {et~al.}(2011)\citenamefont {Rand},
  \citenamefont {Arbesman},\ and\ \citenamefont {Christakis}}]{Rand2011pnas}%
  \BibitemOpen
  \bibfield  {author} {\bibinfo {author} {\bibfnamefont {D.~G.}\ \bibnamefont
  {Rand}}, \bibinfo {author} {\bibfnamefont {S.}~\bibnamefont {Arbesman}}, \
  and\ \bibinfo {author} {\bibfnamefont {N.~A.}\ \bibnamefont {Christakis}},\
  }\href {\doibase 10.1073/pnas.1108243108} {\bibfield  {journal} {\bibinfo
  {journal} {Proceedings of the National Academy of Sciences}\ }\textbf
  {\bibinfo {volume} {108}},\ \bibinfo {pages} {19193} (\bibinfo {year}
  {2011})}\BibitemShut {NoStop}%
\bibitem [{\citenamefont {Lee}\ \emph {et~al.}(2011)\citenamefont {Lee},
  \citenamefont {Holme},\ and\ \citenamefont {Wu}}]{Lee2011prl}%
  \BibitemOpen
  \bibfield  {author} {\bibinfo {author} {\bibfnamefont {S.}~\bibnamefont
  {Lee}}, \bibinfo {author} {\bibfnamefont {P.}~\bibnamefont {Holme}}, \ and\
  \bibinfo {author} {\bibfnamefont {Z.-X.}\ \bibnamefont {Wu}},\ }\href
  {\doibase 10.1103/PhysRevLett.106.028702} {\bibfield  {journal} {\bibinfo
  {journal} {Phys. Rev. Lett.}\ }\textbf {\bibinfo {volume} {106}},\ \bibinfo
  {pages} {028702} (\bibinfo {year} {2011})}\BibitemShut {NoStop}%
\bibitem [{\citenamefont {Hauert}\ and\ \citenamefont
  {Szab\'{o}}(2005)}]{Hauert2005ajp}%
  \BibitemOpen
  \bibfield  {author} {\bibinfo {author} {\bibfnamefont {C.}~\bibnamefont
  {Hauert}}\ and\ \bibinfo {author} {\bibfnamefont {G.}~\bibnamefont
  {Szab\'{o}}},\ }\href {\doibase 10.1119/1.1848514} {\bibfield  {journal}
  {\bibinfo  {journal} {American Journal of Physics}\ }\textbf {\bibinfo
  {volume} {73}},\ \bibinfo {pages} {405} (\bibinfo {year} {2005})}\BibitemShut
  {NoStop}%
\bibitem [{\citenamefont {Doebeli}\ and\ \citenamefont
  {Hauert}(2005)}]{Doebeli2005el}%
  \BibitemOpen
  \bibfield  {author} {\bibinfo {author} {\bibfnamefont {M.}~\bibnamefont
  {Doebeli}}\ and\ \bibinfo {author} {\bibfnamefont {C.}~\bibnamefont
  {Hauert}},\ }\href {\doibase 10.1111/j.1461-0248.2005.00773.x} {\bibfield
  {journal} {\bibinfo  {journal} {Ecology Letters}\ }\textbf {\bibinfo {volume}
  {8}},\ \bibinfo {pages} {748} (\bibinfo {year} {2005})}\BibitemShut {NoStop}%
\bibitem [{\citenamefont {Szab\'{o}}\ and\ \citenamefont
  {F\'{a}th}(2007)}]{Szabo2007pr}%
  \BibitemOpen
  \bibfield  {author} {\bibinfo {author} {\bibfnamefont {G.}~\bibnamefont
  {Szab\'{o}}}\ and\ \bibinfo {author} {\bibfnamefont {G.}~\bibnamefont
  {F\'{a}th}},\ }\href {\doibase
  http://dx.doi.org/10.1016/j.physrep.2007.04.004} {\bibfield  {journal}
  {\bibinfo  {journal} {Physics Reports}\ }\textbf {\bibinfo {volume} {446}},\
  \bibinfo {pages} {97 } (\bibinfo {year} {2007})}\BibitemShut {NoStop}%
\bibitem [{\citenamefont {Roca}\ \emph {et~al.}(2009)\citenamefont {Roca},
  \citenamefont {Cuesta},\ and\ \citenamefont {S\'{a}nchez}}]{Roca2009plr}%
  \BibitemOpen
  \bibfield  {author} {\bibinfo {author} {\bibfnamefont {C.~P.}\ \bibnamefont
  {Roca}}, \bibinfo {author} {\bibfnamefont {J.~A.}\ \bibnamefont {Cuesta}}, \
  and\ \bibinfo {author} {\bibfnamefont {A.}~\bibnamefont {S\'{a}nchez}},\
  }\href {\doibase http://dx.doi.org/10.1016/j.plrev.2009.08.001} {\bibfield
  {journal} {\bibinfo  {journal} {Physics of Life Reviews}\ }\textbf {\bibinfo
  {volume} {6}},\ \bibinfo {pages} {208 } (\bibinfo {year} {2009})}\BibitemShut
  {NoStop}%
\bibitem [{\citenamefont {Perc}\ and\ \citenamefont
  {Szolnoki}(2010)}]{Perc2010bio}%
  \BibitemOpen
  \bibfield  {author} {\bibinfo {author} {\bibfnamefont {M.}~\bibnamefont
  {Perc}}\ and\ \bibinfo {author} {\bibfnamefont {A.}~\bibnamefont
  {Szolnoki}},\ }\href {\doibase
  http://dx.doi.org/10.1016/j.biosystems.2009.10.003} {\bibfield  {journal}
  {\bibinfo  {journal} {Biosystems}\ }\textbf {\bibinfo {volume} {99}},\
  \bibinfo {pages} {109 } (\bibinfo {year} {2010})}\BibitemShut {NoStop}%
\bibitem [{\citenamefont {Nowak}(2012)}]{Nowak2012jtb}%
  \BibitemOpen
  \bibfield  {author} {\bibinfo {author} {\bibfnamefont {M.~A.}\ \bibnamefont
  {Nowak}},\ }\href {\doibase http://dx.doi.org/10.1016/j.jtbi.2012.01.014}
  {\bibfield  {journal} {\bibinfo  {journal} {Journal of Theoretical Biology}\
  }\textbf {\bibinfo {volume} {299}},\ \bibinfo {pages} {1 } (\bibinfo {year}
  {2012})}\BibitemShut {NoStop}%
\bibitem [{\citenamefont {Perc}\ \emph {et~al.}(2013)\citenamefont {Perc},
  \citenamefont {G\'{o}mez-Garde\~{n}es}, \citenamefont {Szolnoki},
  \citenamefont {Flor\'{i}a},\ and\ \citenamefont {Moreno}}]{Perc2013jrsi}%
  \BibitemOpen
  \bibfield  {author} {\bibinfo {author} {\bibfnamefont {M.}~\bibnamefont
  {Perc}}, \bibinfo {author} {\bibfnamefont {J.}~\bibnamefont
  {G\'{o}mez-Garde\~{n}es}}, \bibinfo {author} {\bibfnamefont {A.}~\bibnamefont
  {Szolnoki}}, \bibinfo {author} {\bibfnamefont {L.~M.}\ \bibnamefont
  {Flor\'{i}a}}, \ and\ \bibinfo {author} {\bibfnamefont {Y.}~\bibnamefont
  {Moreno}},\ }\href {\doibase 10.1098/rsif.2012.0997} {\bibfield  {journal}
  {\bibinfo  {journal} {Journal of The Royal Society Interface}\ }\textbf
  {\bibinfo {volume} {10}},\ \bibinfo {pages} {1} (\bibinfo {year}
  {2013})}\BibitemShut {NoStop}%
\bibitem [{\citenamefont {Dawkins}(1976)}]{Dawkins1976book}%
  \BibitemOpen
  \bibfield  {author} {\bibinfo {author} {\bibfnamefont {R.}~\bibnamefont
  {Dawkins}},\ }\href@noop {} {\emph {\bibinfo {title} {The Selfish Gene}}}\
  (\bibinfo  {publisher} {Oxford University Press, New York},\ \bibinfo {year}
  {1976})\BibitemShut {NoStop}%
\bibitem [{\citenamefont {Hamilton}(1963)}]{Hamilton1963an}%
  \BibitemOpen
  \bibfield  {author} {\bibinfo {author} {\bibfnamefont {W.~D.}\ \bibnamefont
  {Hamilton}},\ }\href {http://www.jstor.org/stable/2458473} {\bibfield
  {journal} {\bibinfo  {journal} {Am. Nat.}\ }\textbf {\bibinfo {volume}
  {97}},\ \bibinfo {pages} {354} (\bibinfo {year} {1963})}\BibitemShut
  {NoStop}%
\bibitem [{\citenamefont {Antal}\ \emph {et~al.}(2009)\citenamefont {Antal},
  \citenamefont {Ohtsuki}, \citenamefont {Wakeley}, \citenamefont {Taylor},\
  and\ \citenamefont {Nowak}}]{Antal2009pnas}%
  \BibitemOpen
  \bibfield  {author} {\bibinfo {author} {\bibfnamefont {T.}~\bibnamefont
  {Antal}}, \bibinfo {author} {\bibfnamefont {H.}~\bibnamefont {Ohtsuki}},
  \bibinfo {author} {\bibfnamefont {J.}~\bibnamefont {Wakeley}}, \bibinfo
  {author} {\bibfnamefont {P.~D.}\ \bibnamefont {Taylor}}, \ and\ \bibinfo
  {author} {\bibfnamefont {M.~A.}\ \bibnamefont {Nowak}},\ }\href {\doibase
  10.1073/pnas.0902528106} {\bibfield  {journal} {\bibinfo  {journal}
  {Proceedings of the National Academy of Sciences}\ }\textbf {\bibinfo
  {volume} {106}},\ \bibinfo {pages} {8401} (\bibinfo {year}
  {2009})}\BibitemShut {NoStop}%
\bibitem [{\citenamefont {Camerer}(2003)}]{Camerer2003book}%
  \BibitemOpen
  \bibfield  {author} {\bibinfo {author} {\bibfnamefont {C.}~\bibnamefont
  {Camerer}},\ }\href@noop {} {\emph {\bibinfo {title} {Behavioral Game
  Theory--Experiments in Strategic Interaction}}}\ (\bibinfo  {publisher}
  {Princeton University Press, Princeton, NJ},\ \bibinfo {year}
  {2003})\BibitemShut {NoStop}%
\bibitem [{\citenamefont {Fehr}\ and\ \citenamefont
  {Fischbacher}(2003)}]{Fehr2003nature}%
  \BibitemOpen
  \bibfield  {author} {\bibinfo {author} {\bibfnamefont {E.}~\bibnamefont
  {Fehr}}\ and\ \bibinfo {author} {\bibfnamefont {U.}~\bibnamefont
  {Fischbacher}},\ }\href {\doibase 10.1038/nature02043} {\bibfield  {journal}
  {\bibinfo  {journal} {Nature}\ }\textbf {\bibinfo {volume} {425}},\ \bibinfo
  {pages} {785} (\bibinfo {year} {2003})}\BibitemShut {NoStop}%
\bibitem [{\citenamefont {Szab\'o}\ and\ \citenamefont
  {Szolnoki}(2012)}]{Szabo2012jtb}%
  \BibitemOpen
  \bibfield  {author} {\bibinfo {author} {\bibfnamefont {G.}~\bibnamefont
  {Szab\'o}}\ and\ \bibinfo {author} {\bibfnamefont {A.}~\bibnamefont
  {Szolnoki}},\ }\href {\doibase http://dx.doi.org/10.1016/j.jtbi.2011.03.015}
  {\bibfield  {journal} {\bibinfo  {journal} {Journal of Theoretical Biology}\
  }\textbf {\bibinfo {volume} {299}},\ \bibinfo {pages} {81 } (\bibinfo {year}
  {2012})}\BibitemShut {NoStop}%
\bibitem [{\citenamefont {Szab\'o}\ \emph {et~al.}(2013)\citenamefont
  {Szab\'o}, \citenamefont {Szolnoki},\ and\ \citenamefont
  {Czak\'o}}]{Szabo2013jtb}%
  \BibitemOpen
  \bibfield  {author} {\bibinfo {author} {\bibfnamefont {G.}~\bibnamefont
  {Szab\'o}}, \bibinfo {author} {\bibfnamefont {A.}~\bibnamefont {Szolnoki}}, \
  and\ \bibinfo {author} {\bibfnamefont {L.}~\bibnamefont {Czak\'o}},\ }\href
  {\doibase http://dx.doi.org/10.1016/j.jtbi.2012.10.014} {\bibfield  {journal}
  {\bibinfo  {journal} {Journal of Theoretical Biology}\ }\textbf {\bibinfo
  {volume} {317}},\ \bibinfo {pages} {126 } (\bibinfo {year}
  {2013})}\BibitemShut {NoStop}%
\bibitem [{\citenamefont {Grund}\ \emph {et~al.}(2013)\citenamefont {Grund},
  \citenamefont {Waloszek},\ and\ \citenamefont {Helbing}}]{Thomas2013sr}%
  \BibitemOpen
  \bibfield  {author} {\bibinfo {author} {\bibfnamefont {T.}~\bibnamefont
  {Grund}}, \bibinfo {author} {\bibfnamefont {C.}~\bibnamefont {Waloszek}}, \
  and\ \bibinfo {author} {\bibfnamefont {D.}~\bibnamefont {Helbing}},\ }\href
  {\doibase 10.1038/srep01480} {\bibfield  {journal} {\bibinfo  {journal} {Sci.
  Rep.}\ }\textbf {\bibinfo {volume} {3}},\ \bibinfo {pages} {1480} (\bibinfo
  {year} {2013})}\BibitemShut {NoStop}%
\bibitem [{\citenamefont {Ak\c{c}ay}\ \emph {et~al.}(2009)\citenamefont
  {Ak\c{c}ay}, \citenamefont {Van~Cleve}, \citenamefont {Feldman},\ and\
  \citenamefont {Roughgarden}}]{Akcay2009pnas}%
  \BibitemOpen
  \bibfield  {author} {\bibinfo {author} {\bibfnamefont {E.}~\bibnamefont
  {Ak\c{c}ay}}, \bibinfo {author} {\bibfnamefont {J.}~\bibnamefont
  {Van~Cleve}}, \bibinfo {author} {\bibfnamefont {M.~W.}\ \bibnamefont
  {Feldman}}, \ and\ \bibinfo {author} {\bibfnamefont {J.}~\bibnamefont
  {Roughgarden}},\ }\href {\doibase 10.1073/pnas.0904357106} {\bibfield
  {journal} {\bibinfo  {journal} {Proceedings of the National Academy of
  Sciences}\ }\textbf {\bibinfo {volume} {106}},\ \bibinfo {pages} {19061}
  (\bibinfo {year} {2009})}\BibitemShut {NoStop}%
\bibitem [{\citenamefont {Taylor}\ and\ \citenamefont
  {Nowak}(2007)}]{Taylor2007ev}%
  \BibitemOpen
  \bibfield  {author} {\bibinfo {author} {\bibfnamefont {C.}~\bibnamefont
  {Taylor}}\ and\ \bibinfo {author} {\bibfnamefont {M.~A.}\ \bibnamefont
  {Nowak}},\ }\href {\doibase 10.1111/j.1558-5646.2007.00196.x} {\bibfield
  {journal} {\bibinfo  {journal} {Evolution}\ }\textbf {\bibinfo {volume}
  {61}},\ \bibinfo {pages} {2281} (\bibinfo {year} {2007})}\BibitemShut
  {NoStop}%
\bibitem [{\citenamefont {Blume}(1993)}]{Blume1993geb}%
  \BibitemOpen
  \bibfield  {author} {\bibinfo {author} {\bibfnamefont {L.~E.}\ \bibnamefont
  {Blume}},\ }\href {http://dx.doi.org/10.1006/game.1993.1023} {\bibfield
  {journal} {\bibinfo  {journal} {Games and Economic Behavior}\ }\textbf
  {\bibinfo {volume} {5}},\ \bibinfo {pages} {387 } (\bibinfo {year}
  {1993})}\BibitemShut {NoStop}%
\bibitem [{abc()}]{abc}%
  \BibitemOpen
  \href@noop {} {}\bibinfo {note} {See Supplemental Material for simulation
  results of our model defined on a square lattice as well as on a random
  regular graph with different strategy-updating manners and/or with different
  strategy-updating
  rules~\cite{Blume1993geb,Szabo1998pre,Szabo2007pr}.}\BibitemShut {Stop}%
\bibitem [{\citenamefont {Chen}\ \emph {et~al.}(2008)\citenamefont {Chen},
  \citenamefont {Fu},\ and\ \citenamefont {Wang}}]{Chen2008preb}%
  \BibitemOpen
  \bibfield  {author} {\bibinfo {author} {\bibfnamefont {X.}~\bibnamefont
  {Chen}}, \bibinfo {author} {\bibfnamefont {F.}~\bibnamefont {Fu}}, \ and\
  \bibinfo {author} {\bibfnamefont {L.}~\bibnamefont {Wang}},\ }\href {\doibase
  10.1103/PhysRevE.78.051120} {\bibfield  {journal} {\bibinfo  {journal} {Phys.
  Rev. E}\ }\textbf {\bibinfo {volume} {78}},\ \bibinfo {pages} {051120}
  (\bibinfo {year} {2008})}\BibitemShut {NoStop}%
\bibitem [{\citenamefont {Donaldson}\ and\ \citenamefont
  {Preston}(1995)}]{Thomas1995}%
  \BibitemOpen
  \bibfield  {author} {\bibinfo {author} {\bibfnamefont {T.}~\bibnamefont
  {Donaldson}}\ and\ \bibinfo {author} {\bibfnamefont {L.~E.}\ \bibnamefont
  {Preston}},\ }\href {http://amr.aom.org/content/20/1/65.abstract} {\bibfield
  {journal} {\bibinfo  {journal} {The Academy of Management Review}\ }\textbf
  {\bibinfo {volume} {20}},\ \bibinfo {pages} {65} (\bibinfo {year}
  {1995})}\BibitemShut {NoStop}%
\end{thebibliography}%

\newpage
\begin{widetext}


\section*{Supplementary material}

# Supplemental Material

## Social dilemma alleviated by sharing the gains with immediate neighbors

In this Supplementary Section we present the simulation results of our model defined on a random regular graph where the strategy-updating of the individuals is implemented in an asynchronous manner (**FigS1**); defined on a square lattice as well as on a random regular graph where an alternative random pairwise comparison rule [1-3] is used for the strategy-updating (**FigS2**); defined on a square lattice as well as on a random regular graph where the strategy-updating is implemented in an asynchronous way (**FigS3**). Other simulation settings are the same as those explicated in the main text.

As shown below, whether our model is carried out on a square lattice or on a random regular graph, whether the strategy-updating is implemented in a synchronous or an asynchronous manner, and whether the alternative Fermi-function like rule is used for the strategy updating, will yield qualitatively similar results as those illustrated in the main text. Hence the two main findings in our work, (i) *cooperation can be promoted substantially as the players are becoming more and more related with each other* and (ii) *cooperation can be more readily established in the context of the PDG than that in the SG with appropriate parameters*, are robust to different topological structures of the interaction network, different strategy-updating functions, and different strategy-updating manners.

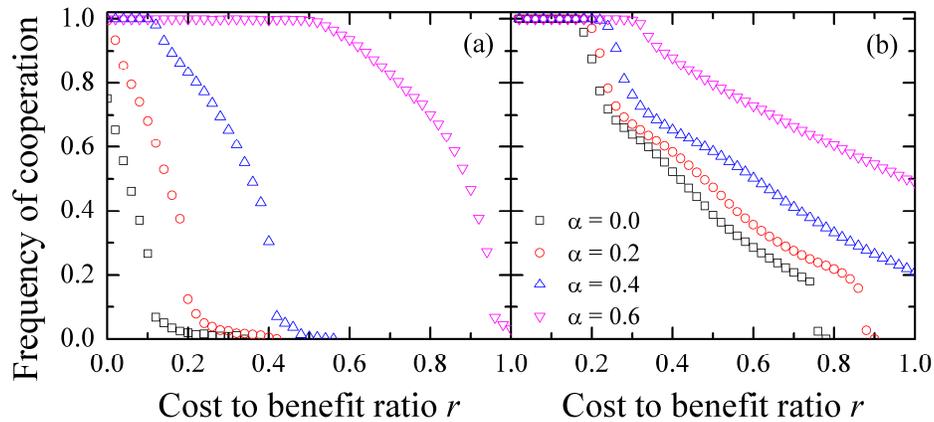

**FigS1:** Frequency of cooperation as a function of the cost-to-benefit ratio (a) $r = c/(b - c)$ in the PDG and (b) $r = c/(2b - c)$ in the SG for different values of $\alpha$, as denoted in the figure legend. The players are located on the sites of a random regular graph of size $N = 65536$, where each player has exactly four interaction neighbors. The strategy-updating is implemented in a synchronous manner, just the same as what we did in the main text.

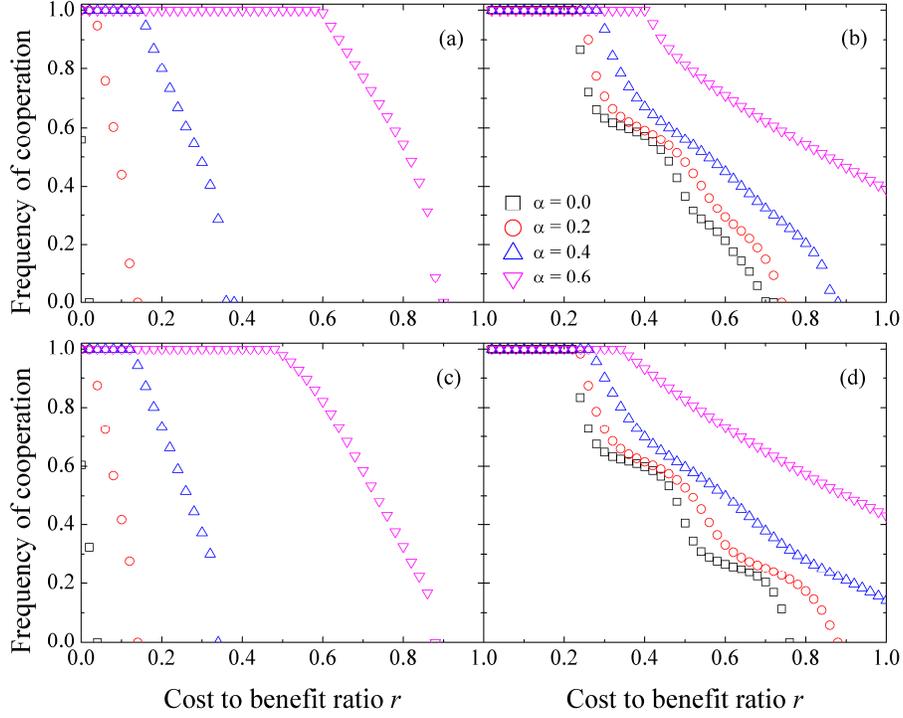

**FigS2:** Frequency of cooperation as a function of the cost-to-benefit ratio $r = c/(b - c)$ in the PDG [(a) and (c)], and $r = c/(2b - c)$ in the SG [(b) and (d)] for different values of $\alpha$. (a) and (b) are for the case where the players are located on the sites of a square lattice of size $N = 256*256$. (c) and (d) are for the case where the players are located on the sites of a random regular graph of size $N = 65536$. The strategy-updating is implemented by the random pairwise comparsion rule [1-3] $W(i \leftarrow j) = \dfrac{1}{1+\exp[(G_i - G_j)/\kappa]}$ with a synchronous manner, where $W(i \leftarrow j)$ is the probability that the individual $i$ will adjust to the strategy of the randomly selected neighbor $j$, and $\kappa$ is set to 0.25 representing the strength of noise.

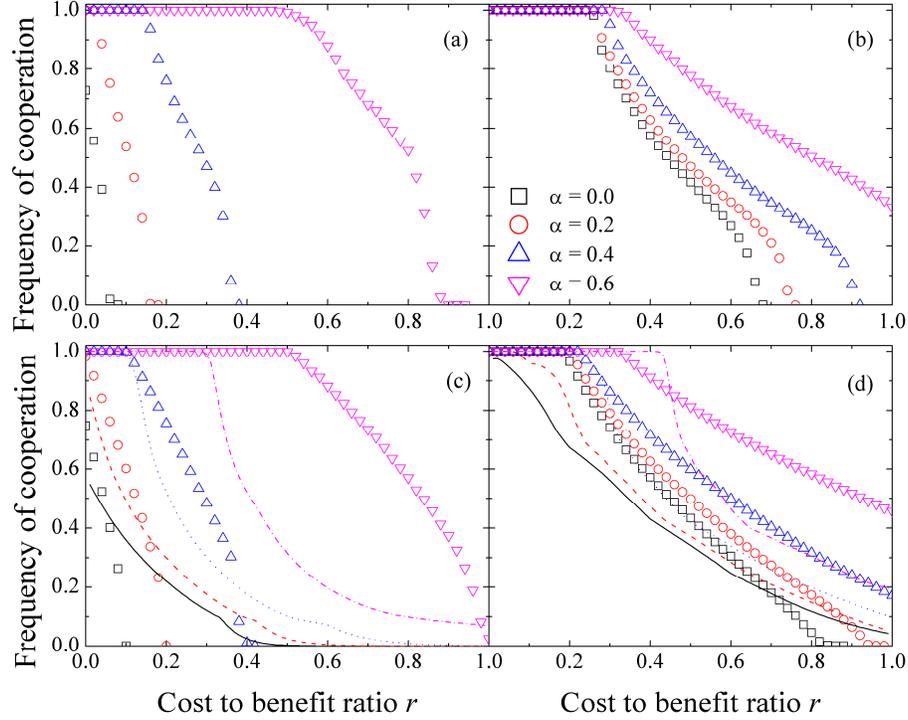

**FigS3:** Frequency of cooperation as a function of the cost-to-benefit ratio $r = c/(b - c)$ in the PDG [(a) and (c)], and $r = c/(2b - c)$ in the SG [(b) and (d)] for different values of $\alpha$. (a) and (b) are for the case where the players are located on the sites of a square lattice of size $N = 256*256$. (c) and (d) are for the case where the players are located on the sites of a random regular graph of size $N = 65536$. The strategy-updating is implemented by using the Eq.(2) in the main text in an asynchronous manner. The curves in (c) and (d) correspond to the predictions by the extended pair approximation approach, elaborated in the Appendix in the main text.

**References:**
1. L. E. Blume, Games Econ. Behav. **5**, 387 (1993).
2. G. Szabó and C. Tőke, Phys. Rev. E **58**, 69 (1998).
3. C. Hauert and G. Szabó, Am. J. Phys. **73**, 405 (2005).


\end{widetext}
\end{document}